\documentclass[12pt]{article}

\usepackage{amsmath}
\usepackage{amsfonts}
\usepackage{amssymb}
\usepackage{enumerate}
\usepackage{bm}
\usepackage{graphicx}

\usepackage{amsthm} 
\newtheorem{theorem}{Theorem}[section]

\newtheorem{lemma}[theorem]{Lemma}

\newtheorem{definition}[theorem]{Definition}

\newcommand{\R}{\mathbb{R}}
\newcommand{\Z}{\mathbb{Z}}

\newcommand{\C}{\mathbb{C}}

\newcommand{\s}{\sigma}
\newcommand{\w}{\omega}

\newcommand{\bu}{\bm{u}}
\newcommand{\bv}{\bm{v}}

\title{A genus six cyclic tetragonal reduction of the Benney equations}

\author{M England$^1$ and J Gibbons$^2$}

\begin{document}

\maketitle

\begin{footnotesize}
\noindent $^1$ Department of Mathematics and the Maxwell Institute for Mathematical Sciences,
School of Mathematical and Computer Sciences (MACS),
Heriot-Watt University,
Riccarton, Edinburgh,
EH14 4AS, UK \quad \textit{M.England@ma.hw.ac.uk} \\
$^2$ Department of Mathematics,
Imperial College London,
South Kensington Campus,
London,
SW7 2AZ, UK \quad \textit{j.gibbons@imperial.ac.uk}
\end{footnotesize}

\vspace*{0.1in}

\begin{abstract}
A reduction of Benney's equations is constructed corresponding to 
Schwartz-Christoffel maps associated with a family of genus six cyclic tetragonal curves.
The mapping function, a second kind Abelian integral on the associated Riemann surface,
is constructed explicitly as a rational expression in derivatives of the
Kleinian $\sigma$-function of the curve.
\end{abstract}

% \ams{14H40, 14H42, 14H70, 14H51, 33F10}
% \pacs{02.30.Ik, 02.30.Gp, 02.70.Wz}

\section{Introduction} \label{SEC_intro}

In \cite{GibTsa1, GibTsa2} it was shown that Benney's equations
$$A^n_t = A^{n+1}_x +n A^{n-1} A^0_x,\qquad n\ge 0,$$
admit {\em reductions} in which only finitely many $N$ of the moments $A^n$
are independent, and that a large class of such reductions may be parametrised
by conformal maps from the upper half $p$-plane to a slit domain - the upper
half $\lambda$-plane, cut along $N$ nonintersecting Jordan arcs, which have
one fixed end point on $\Im(\lambda)=0$, and whose other \lq free' end 
is a Riemann invariant of the reduced equations.

A natural subclass of these occurs where these Jordan arcs are straight lines,
leading to a polygonal domain and hence an $N$-parameter 
Schwartz-Christoffel map; an important and tractable subfamily of 
these is the case in which the angles are all rational multiples of $\pi$,
and in this case the mapping is given by an integral of a second kind 
Abelian differential \cite{yg99} on an algebraic curve. 
Such examples have been worked out explicitly, in \cite{yg99}, \cite{bg03},
\cite{bg04} and \cite{bg06}. 
These have looked at elliptic and hyperelliptic curves as well as a 
cyclic trigonal example. It is thus worthwhile to generalise this to 
other algebraic curves.

In all these examples, reductions have been constructed explicitly - both
the integrand and its integral were evaluated using quotients of 
derivatives of the  $\sigma$-function associated with the respective 
curves. These curves are all specific examples 
from the wider class of cyclic $(n,s)$ curves which have equations of the
form: 
\begin{equation} \label{ns_curve}
y^n = x^s + \mu_{s-1} x^{s-1} + \dots +\mu_1 x + \mu_0.
\end{equation}
We suppose that $(n,s)$ are coprime with $n<s$, in which case the 
curves have genus $g = \frac{1}{2}(n-1)(s-1)$, and a unique branch 
point $\infty$ at infinity.  In this paper we consider a reduction associated with a cyclic tetragonal curve, that
is, from the class (1) with $n = 4$. For simplicity we look at the case with $s=5$ here.

\section{Benney's Equations} \label{SEC_BE}

In 1973, Benney considered an approximation for the two-dimensional \\ equations of motion of an incompressible perfect fluid under a gravitational force~\cite{benney}.  He showed that if moments are defined by:
\[ 
A_n (x,t) = \int_{0}^{h}{u^n \, \mathrm{d}y },
\]
where $u(x,y,t)$ is the horizontal fluid velocity and $h(x,t)$ the 
height of the free surface,  the moments $A_n(x,t)$ satisfy an infinite 
set of hydrodynamic type equations
\begin{equation} \label{eq:benneymoment}
 \frac{ \partial A_n}{ \partial t} + \frac{ \partial A_{n+1} }{\partial
 x} + n \, A_{n-1} \, \frac{ \partial A_0}{\partial x} =0  \qquad \qquad
 (n=1,2,\dots),
\end{equation}
now called the Benney moment equations.

Identical moment equations can alternatively be derived from a Vlasov 
equation~\cite{Gib1}, \cite{zakh1}:
\begin{equation} \label{eq:vlasov}
\frac{ \partial f}{\partial t_2} + p \, \frac{ \partial f}{ \partial x}
- \frac{ \partial A_0}{\partial x} \frac{ \partial f}{\partial p} = 0.
\end{equation}
Here $ f= f(x,p,t)$ is a distribution function and the moments are
defined instead by
\[ A_n = \int_{-\infty}^{\infty}{ p^n f} \, \mathrm{d}p.\]
We assume throughout that $f$ is such that all these moments exist.
The equation of motion~(\ref{eq:vlasov}) has the Lie-Poisson structure :
\begin{equation} \label{eq:hamstruct}
\frac{ \partial f}{\partial t} + 
\left\{ f  , \frac{ \delta H}{\delta f} \right\}_{p,x}=0,
\end{equation}
where $ \left\{ \cdot \, , \cdot \right\}_{p,x}$ is the canonical Poisson
bracket. 
Kupershmidt and Manin showed directly that the moment equations are
Hamiltonian~\cite{kupman1},~\cite{kupman2}. 
If we set
$ H= \frac{1}{2} H_2 = \frac{1}{2}( A_2 +A_0^2), \; \mathbf{A}=( A_0,
A_1, \ldots ), $ then
\begin{equation} \label{eq:kupandman}
\frac{ \partial \mathbf{A}}{\partial t} = B \frac{ \partial
H}{\partial \mathbf{A}}
\end{equation}
where the matrix operator $B$ is given by
\[
B\, \!_{n,m}=  n A_{n+m-1} \frac{ \partial}{\partial x}  + 
m \frac{ \partial}{\partial x} \cdot A_{n+m-1}.
\]
This is consistent with~(\ref{eq:hamstruct}) in the sense that if $H$ 
is some function only of the moments, the moment equations resulting 
from~(\ref{eq:hamstruct}) and~(\ref{eq:kupandman}) are identical.

Benney showed in~\cite{benney} that system~(\ref{eq:benneymoment}) has 
infinitely many conserved densities, polynomial in the $A_n.$  
One of the most direct ways
to calculate these is to use generating functions~\cite{kupman1}.
Let $\lambda(x,p,t),$ a formal series in $p,$ be the generating function
of the moments
\begin{equation} \label{eq:lambda1}
 \lambda(x,p,t)= p + \sum_{n=0}^{\infty}{ \frac{ A_n}{p^{n+1}}} 
\end{equation}
and let $p(x, \lambda, t)$ be the inverse series 
\[ p(x, \lambda, t) = \lambda - \sum_{m=0}^{\infty}{ \frac{H_m}{\lambda^{m+1}}}.\]
We note here that if 
$A_n=\int_{-\infty}^{\infty}{p^n f \, \mathrm{d}p} \, $ is
substituted into~(\ref{eq:lambda1}), then this can be understood as the 
asymptotic series, as $ p \to \infty$, of an integral
\begin{equation}
 \lambda= 
 p + \int_{-\infty}^{\infty}{ \frac{f(x,p',t)}{(p-p')} } \,
\mathrm{d}p'. \label{eq:lambda2}
\end{equation}
Here $p'$ runs along the real axis, and we take $\Im(p)>0$.  It follows that
$\lambda(p)$ is analytic in its domain of definition. If $f(p)$ is 
H\"older continuous, the boundary value, on the real 
$p$-axis, of $\lambda(p)$  will itself be H\"older continuous.

Comparing the first derivatives of $\lambda(x,p,t),$  we obtain the PDE
\begin{equation} \label{eq:lambdaderivs}
\frac{\partial \lambda}{\partial t} + p \, \frac{ \partial \lambda}{\partial
x} = \frac{ \partial \lambda}{\partial p} \left( \frac{ \partial p}{ \partial
t} + p \, \frac{ \partial p}{\partial x}  + \frac{ \partial A_0}{\partial
x} \right). 
\end{equation}
If we now hold $p$ constant, this gives
\begin{equation} \label{eq:vlasov2}
\frac{\partial \lambda}{\partial t} + p \, \frac{ \partial \lambda}{\partial
x} - \frac{ \partial A_0}{\partial x} \,\frac{ \partial \lambda}{\partial p}  =0
\end{equation}
which is a Vlasov equation of the same form
as~(\ref{eq:vlasov}).  Thus (\ref{eq:vlasov}) and (\ref{eq:vlasov2}) 
have the same characteristics.  Any function of $\lambda$ and $f$ must 
satisfy the same equation. 

Alternatively, if we hold $\lambda$ constant 
in~(\ref{eq:lambdaderivs}), then we obtain the
conservation equation
\begin{equation} \label{eq:conserv}
\frac{\partial p}{\partial t} + 
\frac{\partial}{\partial x} \left( \frac{1}{2} p^2 + A_0 \right) = 0.
\end{equation}
Substituting the formal series of $p(x,\lambda, t)$ into (\ref{eq:conserv}),
we see that each
$H_n$ is polynomial in the $A_n$ and is a conserved density. 
Any of the $H_n$ could be used as
the Hamiltonian in~(\ref{eq:hamstruct}), and  the resulting flows all commute.
From this we define the Benney hierarchy to be the family of 
evolution equations
\[ \frac{ \partial f}{\partial t_n} +  \left\{ f \, , \, \frac{1}{n} \frac{ \delta
H_n}{\delta f} \right\} = 0. \]
Again,  $\lambda$ satisfies an equation analogous to this,
\[ \frac{ \partial \lambda}{\partial t_n} +  \left\{ \lambda \, , \, \frac{1}{n} \frac{ \delta
H_n}{\delta f} \right\} = 0, \]
so that both $f$ and $\lambda$ are advected along the same characteristics.
These characteristics are flows of a Hamiltonian vector field, with 
Hamiltonians $\frac{ \delta H_n}{\delta f}$   given by the relation: 
\[ \left( \frac{1}{n} \frac{ \delta H_n}{\delta f} \right) = \left(
\frac{\lambda^n}{n} \right)_+\]
where $( \cdot)_+$ denotes the polynomial part of the Laurent expansion.

\subsection{Reductions of the moment equations} \label{SUBSEC_Reduction}
 Suppose that for some family of points,
 $p=\hat{p}_i(x,t), \,  \lambda(\hat{p}_i)=\hat{\lambda}_i(x,t) $, 
we have 
\[ \left. \frac{  \partial \lambda}{\partial p} \right|_{p=\hat{p}_i}=0.\]
Then~(\ref{eq:lambdaderivs}) reduces to: 
\[ \frac{ \partial \hat{\lambda}_i}{\partial t} + \hat{p}_i \, \frac{ \partial
\hat{ \lambda}_i}{\partial x} =0 \]
where 
$\frac{ \partial \hat{\lambda}_i}{\partial t} = \left.
\frac{  \partial \lambda}{\partial t} \right|_{p=\hat{p}_i} 
$ and 
$ \, \frac{ \partial \hat{\lambda}_i}{\partial x} = \left.
\frac{  \partial \lambda}{\partial x} \right|_{p=\hat{p}_i}.$
We say that $\hat{ \lambda}_i$ is a Riemann invariant with 
characteristic speed $ \hat{p}_i.$ We will see that there are families of
functions $\lambda(p)$
which are invariant under the Benney dynamics, and are parametrised by $N$
Riemann invariants $\lambda_i$.

A hydrodynamic type system with $N\ge 3$, independent variables
can not in general be expressed in terms of Riemann invariants.
If such a system does have $N$ Riemann invariants, it is called diagonalisable.
Tsarev showed in~\cite{Tsa1} that if a diagonal hydrodynamic-type
system 
\begin{equation} \label{eq:diagv}
\frac{ \partial \hat{\lambda}_i}{ \partial t} + v_i(\hat{\lambda}) \frac{ \partial
\hat{\lambda}_i}{\partial x} =0  \qquad (i=1,2,\dots,N).
\end{equation}
is semi-Hamiltonian, that is if 
\[ \partial\, \! _j \left( \frac{ \partial\, \!_i v_k}{ v_i -
v_k}  \right)  = 
\partial\, \! _ i \left( \frac{ \partial\, \!_j v_k}{ v_j-v_k}
\right), \qquad i  \neq j \neq k,
\]
for $i,j,k$ distinct, where
\[ 
\partial_k = \frac{ \partial }{ \partial \hat{\lambda}_k},
\]
then it can be solved by the hodograph transformation. Any Hamiltonian
system of hydrodynamic type is semi-Hamiltonian. Given a second
equation of type~(\ref{eq:diagv})
\begin{equation} \label{eq:diagw}
\frac{ \partial \hat{\lambda}_i}{ \partial \tau} + {w}_i(\hat{\lambda}) \frac{ \partial
\hat{\lambda}_i}{\partial x} =0 \qquad ( i=1,2,\dots, N),
\end{equation}
and requiring it to be consistent with~(\ref{eq:diagv}), we find that 
the ${w}_i( \hat{\lambda})$ must satisfy the over-determined linear system 
\begin{equation} 
\frac{ \partial\, \!_k {w}_i}{{w}_i - {w}_k} = 
\frac{ \partial\, \!_k {v}_i}{ {v}_i - {v}_k} , \qquad i \neq k.\label{commute}
\end{equation} 
These equations are consistent provided~(\ref{eq:diagv}) is 
semi-Hamiltonian. 
If the condition~(\ref{commute}) holds, we say that ~(\ref{eq:diagv}) 
and ~(\ref{eq:diagw}) commute. 
In this case a set of equations for the unknowns $\hat{\lambda}_i(x,t)$ 
is given by :
\[{w}_i(\hat{\lambda})={v}_i(\hat{\lambda})\, t+x, \qquad (i=1,2,\dots ,N)
\]
where $t$ and $x$ are the independent variables.
Thus any reduction of this type can be solved in principle.

This generalized hodograph construction cannot easily be applied 
directly to the Benney equations however, as these have infinitely many 
dependent variables. Instead we will now consider families of 
distribution functions $f$, which are
parameterised by finitely many $N$ Riemann invariants 
$\hat{\lambda}_{i} ( x, t).$
We are interested in the case \cite{GibTsa1}, \cite{GibTsa2} 
where the function $\lambda(p,x,t)$ is such
that only $N$ of the moments are independent. Then 
there are $N$ characteristic speeds, assumed real and distinct, and $N$
corresponding Riemann invariants 
$(\hat{p}_i, \hat{\lambda}_i),$ so Benney's equations reduce to a 
diagonal system of hydrodynamic type with finitely many dependent 
variables $\hat{\lambda}_i$,
\begin{equation} \label{eq:diagp}
\frac{ \partial \hat{\lambda}_i}{ \partial t} + 
 \hat{p}_i(\hat{\lambda}) \frac{ \partial \hat{\lambda}_i}{\partial x} 
 =0  \qquad (i=1,2,\dots,N).
\end{equation}
Such a system is called a reduction of Benney's equations.

The construction of a more general family of solutions for equations 
of this type was outlined in~\cite{GibTsa1} and~\cite{GibTsa2}.   
An elementary example is the case where the map $\lambda_+$ takes the
upper half $p$-plane to the upper half $\lambda$-plane with a vertical
slit as follows. This is a Schwarz-Christoffel map:
\[ \lambda_+(x,p,t)= p + \int_{\infty}^{p}{ \frac{ p'-\hat{p}_1}{ \sqrt{
(p'-p_1)(p'-p_2)}} \: \mathrm{d}p'}. \]
If the residue at infinity is set to be zero, then this imposes the condition
$\hat{p}_1 = \frac{1}{2}(p_1+p_2)$ and we get the solution
\begin{eqnarray*}
 \lambda_+(x,p,t) & = \hat{p}_1 + \sqrt{ p^2- (p_1+p_2)\,\! p +p_1p_2} \\
               & = \hat{p}_1 + \sqrt{ (p- \hat{p}_1)^2 + 2 A_0}
\end{eqnarray*}
(from the expansion as $p \to \infty$).
This gives a steadily translating solution of Benney's equations~(\ref{eq:hamstruct})
\[
\frac{ \partial f}{\partial t} + \left\{ f \,  , \, \frac{1}{2} p\, \! ^2 + A_0
\right\}_{p,x}=0.
 \]
The two parameters $p_1$ and $p_2$ are not independent, as for consistency
their sum must be a constant. Hence only the end point of the slit in the
$\lambda$-plane is variable. This is the Riemann invariant. 

This construction can be generalized, \cite{GibTsa2}, mapping the upper half $p$-plane to the upper half $\lambda$-plane with $N$ curvilinear slits.
However, in this paper we are specifically interested in {\em straight} slits -
these mappings are all of Schwartz-Christoffel type.

\subsection{Schwartz-Christoffel reductions} \label{SUBSEC_SC}
The case of a polygonal $N$-slit domain is of particular interest. The real
$p$-axis has $M$ vertices $u_{j}$ marked on it; the preimage in the $p$-plane
of each slit runs from a vertex $\hat{p}_j$, to a point $\hat{v}_i$, the
preimage of the end of the slit, and then to another vertex 
$\hat{p}_{j+1}$. The angle $\pi$ in the $p$-plane at $\hat{p}_j$ is 
mapped to an angle $\alpha_j \pi$ at the image point. The internal angle at the end of each slit is $2\pi$.

The mapping function is then given, up to a constant of integration, by
$$ \lambda =  \int^p \left[   
\frac{\prod_{i=1}^N(p-\hat{v}_i)}{\prod_{j=1}^{2N} (p-\hat{p}_j)^{1-\alpha_j}} 
\right] \mathrm{d}p.$$
If the integrand is to converge to $1$ as $p\rightarrow \infty$, we require
$\sum_{j=1}^{2N} \alpha_j = N$, while to avoid a logarithmic singularity,
we further impose $\sum_{j=1}^{M} \alpha_j \hat{p}_j = \sum_{i=1}^N \hat{v}_i$.
We then define $\lambda$ more precisely as
$$ \lambda = p+ \int^p_{-\infty} \left[   
\frac{\prod_{i=1}^N(p-\hat{v}_i)}{\prod_{j=1}^{M} (p-\hat{p}_j)^{1-\alpha_j}}
-1 \right] \mathrm{d}p.$$
Other constraints are imposed by requiring the vertices $\hat{p}_j$ to 
map to points $\lambda_i^{0}$, the fixed base points of the slits; 
there remain $N$ independent parameters, which can
be taken to be the movable end points of the slits $\lambda_i(x,t)$. These
satisfy the equations of motion
\[ \frac{\partial \lambda_i}{\partial t} + 
\hat{v}_i  \frac{\partial \lambda_i}{\partial x} =0. \]
To understand and to solve these equations, it is necessary to understand
the dependence of the $\hat{v}_i$ on the Riemann invariants 
$(\lambda_1, \ldots,\lambda_N)$ 
- we thus need to evaluate the Schwartz-Christoffel integral explicitly.

The most tractable cases are where all the $\alpha_j$ are rational, so that
the integrand becomes a meromorphic second kind differential on some algebraic
curve. In this case the only singularity is as $p \rightarrow \infty$, where
the integrand has a double pole with no residue, and the integral thus 
has a simple pole.
For specific families of curves, as in \cite{yg99}, \cite{bg03},
\cite{bg04} and \cite{bg06}, this integral has been worked out explicitly.
In each case, these mappings were found as rational functions of derivatives
of the Kleinian $\sigma$-function of the associated curve.

\section{A tetragonal reduction} \label{SEC_TR}

\begin{figure}[h] 
\begin{center}
\includegraphics*[width=12cm]{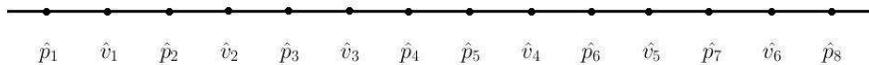}
\caption{The domain $\mathcal{P}$ within the $p$-plane. \label{fig_pplane}}
\end{center}
\end{figure}

We will consider reductions that allow us to work on a tetragonal surface;
these have not been considered before.  Such reductions will require 
two or more sets of straight slits, making angles of 
$\pi/4, \pi/2, 3\pi/4$ to the horizontal.    
Define $\mathcal{P}$ to be the upper half $p$-plane with 14 points 
marked on the real axis, as in Figure \ref{fig_pplane}.  

\begin{figure}[t] 
\begin{center}
\includegraphics*[width=12cm]{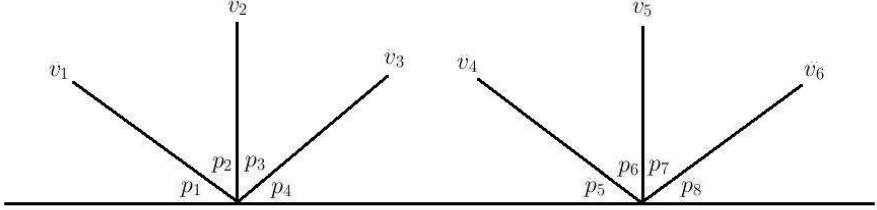}
\caption{The domain $\mathcal{L}'$ within the $\lambda$-plane. \label{fig_lplane}}
\end{center}
\end{figure}

\noindent These points satisfy
\[
\hat{p}_1 < \hat{v}_1 < \hat{p}_2 < \hat{v}_2 < \hat{p}_3 < \hat{v}_3 < \hat{p}_4 < 
\hat{p}_5 < \hat{v}_4 < \hat{p_6} < \hat{v_5} < \hat{p_7} < \hat{v_6} < \hat{p}_8.
\]
Then define the domain $\mathcal{L}'$ as the upper-half $\lambda$-plane with two triplets of slits, as described above. 
We let the first trio of slits radiate from the fixed point $p_1$, with the end points of these three slits labelled $v_1,v_2$ and $v_3$ respectively. Similarly, let the second trio of slits radiate from $p_5$ and have end points $v_4,v_5,v_6$.  Finally impose the conditions that 
\begin{eqnarray*} 
\lambda(\hat{p}) &= p, \qquad p_1 = p_2 = p_3 = p_4, \\
\lambda(\hat{v}) &= v, \qquad p_5 = p_6 = p_7 = p_8.
\end{eqnarray*}
We then see that $\mathcal{L}'$ is the slit domain as shown in Figure \ref{fig_lplane}, and the mapping 
$\lambda: \mathcal{P} \to \mathcal{L}'$ can be given in Schwartz Christoffel form by 
\begin{equation}
\lambda(p) = p + \int_{\infty}^{p} \big[ \varphi(p') - 1 \big] dp' \label{TheMapping}
\end{equation}
where 
\begin{equation}
\varphi(p) = \frac{ \prod_{i=1}^6 (p - \hat{v}_i) }{ \big[ \prod_{i-1}^8 (p-\hat{p}_i) \big]^{\frac{3}{4}}} = 
\frac{ \prod_{i=1}^6 (p - \hat{v}_i) }{ y^3} \label{integrand1}
\end{equation}
where
\begin{equation}
y^4 = \prod_{i=1}^8 (p - \hat{p}_i). \label{48curve}
\end{equation}

\noindent Note that we require the following zero residue property:
\begin{equation}
\lim_{p \to \infty} \varphi(p) \sim 1 + O\left(\frac{1}{p^2} \right). \label{zr1}
\end{equation}

\begin{figure}[t] 
\begin{center}
\includegraphics*[width=10.5cm]{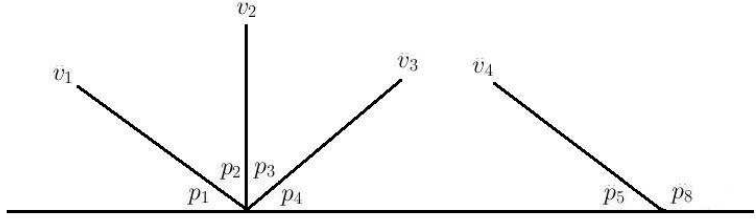}
\caption{The domain $\mathcal{L}$ within the $\lambda$-plane. \label{fig_lplane_r}}
\end{center}
\end{figure}

This mapping would lead us to consider the Riemann surface given by 
points $(p,y)$ that satisfy (\ref{48curve}).  However, we wish to 
consider the simplest possible tetragonal surface (one with only six 
branch points) and so we collapse two of the slits, (the final two by 
choice).  This simplifies our $\lambda$-plane to $\mathcal{L}$, given 
in Figure \ref{fig_lplane_r}.

Further, as in the trigonal case, the analysis of this surface is eased 
if we put it into canonical form, by mapping one of the branch points 
($p_8$ by choice), to infinity.  So we use the following invertible 
rational map to perform these simplifications on our curve and integrand.
\begin{align} 
\hat{p}_6 &= \hat{p}_8, \quad \hat{p}_7 = \hat{p}_8, \quad \hat{v}_5 = \hat{p}_8, \quad 
\hat{v}_6 = \hat{p}_8, \nonumber \\
p &= \hat{p}_8 - (1/t),  \qquad \hat{p}_i = \hat{p}_8 - 
\textstyle  \frac{1}{T_i}, \quad i=1,\dots,5 \label{canonical_map}  \\
y &=   \frac{sk}{t^2} \hspace*{0.2in} \mbox{where} \hspace*{0.2in} 
k^4 =  - \prod_{i=1}^5 (\hat{p}_8 - \hat{p}_i) = -\prod_{i=1}^5 \frac{1}{T_i}   
\nonumber 
\end{align}
If we perform the mapping (\ref{canonical_map}) on the curve (\ref{48curve}) we obtain
\begin{align}
\frac{s^4k^4}{t^8} &= 
\left[ \prod_{i=1}^5 \left(\hat{p}_8 - \frac{1}{t} - \hat{p}_8 
+ \frac{1}{T_i}\right) \right]\left(\hat{p}_8 - \frac{1}{t} - \hat{p}_8\right)^3 
= -\frac{1}{t^3} \prod_{i=1}^5 \left(\frac{1}{T_i} 
- \frac{1}{t} \right)  \nonumber\\ 
&= -\frac{1}{t^3} \left[ \prod_{i=1}^5 
\left(t - T_i \right)\frac{1}{tT_i} \right]
= \left[ \prod_{i=1}^5 \left(t - T_i \right) \right] 
\cdot \left(\frac{1}{t^8}\right) \cdot 
(-1)\left[ \prod_{i=1}^5 \frac{1}{T_i} \right]. \nonumber 
\end{align}
This simplifies to give the following canonical form of (\ref{48curve}).
\begin{align}
s^4 &= \prod_{i=1}^5 \left(t - T_i \right) = 
t^5 + \mu_4t^4 + \mu_3 t^3 + 
\mu_2t^2 + \mu_1t + \mu_0,  \label{45curve_bp}
\end{align}
for constants $\mu_0,\dots,\mu_4$.  Let $C$ denote the Riemann 
surface defined by (\ref{45curve_bp}).  

We now consider $\lambda(p)$ as mapping $\mathcal{P} \to \mathcal{L}$ by performing 
(\ref{canonical_map}) on the integrand  (\ref{integrand1}).
\begin{align}
\varphi(p) dp &= 
\left( \frac{t^6}{s^3k^3} \right) 
\left[ \prod_{i=1}^4 \left(\hat{p}_8 - \frac{1}{t} - \hat{v}_i\right) \right] 
\left(\hat{p}_8 - \frac{1}{t} - \hat{p}_8 \right)^2  
\left( -\frac{1}{t^2}dt \right) \nonumber \\
&= \left( \frac{t^6}{s^3k^3} \right) 
\left[ \prod_{i=1}^4 \left( (\hat{p}_8 - \hat{v}_i)t -1\right) \right]
\left[\frac{1}{t^4}\right] \left( \frac{1}{t} \right)^2  
\left( -\frac{1}{t^2}dt \right) \nonumber \\
&= K \left[ A_4t^4 + A_3t^3 + A_2t^2 + A_1t + 1 \right] 
\frac{1}{t^2}   \frac{dt}{4s^3}  \equiv \varphi(t) dt,
\label{integrand2}
\end{align}
where $K = -4/k^3$ and $A_1,\dots,A_4$ are constants.  
We will evaluate this integrand using algebraic functions defined upon $C$.
% the Riemann surface (\ref{45curve_bp}).

\section{Properties of the tetragonal surface $\bm{C}$} \label{SEC_PropertiesC}

The Riemann surface $C$ is defined by (\ref{45curve_bp}), the cyclic 
tetragonal curve of genus six.  This is also referred to as the 
\emph{cyclic (4,5)-curve}.   The surface is constructed from four 
sheets of the complex plane, with branch points of order 4 at 
$T_1,\dots,T_5,T_6=\infty$, a local coordinate at $t=\infty$ given by 
$\xi=t^{-1/4}$ and branch cuts along the intervals
\[
[T_1,T_2], \quad [T_2,T_3], \quad [T_3,T_4], \quad [T_5,\infty].
\]
This surface was recently considered in \cite{MYPAPER}, where the aim 
was to generalise the theory of the Weierstrass $\wp$-function to 
Abelian functions associated with (\ref{45curve_bp}).  We will give the 
essential properties of the surface here, but refer the reader to 
\cite{MYPAPER} for some of the details and proofs.

We start by noting that there are a set of \emph{Sato weights} 
associated to the surface, which render all equations within the theory 
homogeneous.  The weights of the variables $t,s$ and the curve 
constants can be determined up to a constant factor, from the curve 
equation. We may set the weights as below. 
\begin{center}
\begin{tabular}{|c|c|c|c|c|c|c|c|c|c|c|c|c|}\hline
                & $t$  & $s$   &  $\mu_4$ & $\mu_3$ & $\mu_2$ & $\mu_1$ & $\mu_0$    \\ \hline
\textbf{Weight} & $-4$ & $-5$  &  $-4$        & $-8$        & $-12$       & $-16$       & $-20$          \\ \hline
\end{tabular}
\end{center}
The weights of other variables and function in the theory are derived 
uniquely from these, with all other constants are assigned zero weight. 

\noindent We define a basis of holomorphic differentials upon $C$ by
\begin{align}
\bm{du} &= (du_1, \dots ,du_6), \qquad 
du_i(t,s) = \frac{g_i(t,s)}{4s^3}dt, \nonumber \\
&\mbox{where} \qquad 
\begin{array}{lllll}
g_1(t,s) = 1,   & \quad & g_2(t,s) = t, & \quad & g_3(t,s) = s, \\
g_4(t,s) = t^2, & \quad & g_5(t,s) =ts, & \quad & g_6(t,s) = s^2. \label{holodiff}
\end{array}
\end{align}
We can use the local parameter $\xi$ to express these as series, 
\begin{align} \label{duxi}
\begin{array}{lll}
du_1 = -\xi^{10} + O(\xi^{11})  d\xi   & \quad &
du_4 = \textstyle -\xi^{2} + \frac{3}{4}\mu_4\xi^{6} + O(\xi^7) d\xi \\
du_2 = -\xi^{6} + O(\xi^7) d\xi & \quad &
du_5 = \textstyle -\xi^{1} + \frac{1}{2}\mu_4\xi^5 + O(\xi^6) d\xi \\
du_3 = -\xi^{5} + O(\xi^6) d\xi & \quad & 
du_6 = \textstyle -1 + \frac{1}{4}\mu_4\xi^4 + O(\xi^5) d\xi. 
\end{array}
\end{align}
We know from the general theory that any point $\bu \in \C^6$ can be expressed as 
\begin{align*}
\bu &= (u_1,u_2,u_3,u_4,u_5,u_6) = \sum_{i=1}^6 \int_{\infty}^{P_i} \bm{du},
\end{align*}
where the $P_i$ are six variable points upon $C$. Integrating (\ref{duxi}) gives 
\begin{align}
\begin{array}{lll}
u_1 = -\frac{1}{11}\xi^{11} + O(\xi^{15}) & u_3= -\frac{1}{6}\xi^6 + O(\xi^{10}) & u_5 = -\frac{1}{2}\xi^2 + O(\xi^6) \\
u_2 = -\frac{1}{7}\xi^{7} + O(\xi^{11})   & u_4= -\frac{1}{3}\xi^3 + O(\xi^7)    & u_6 = -\xi + O(\xi^5).
\end{array} \label{uxi}
\end{align}
from which we can conclude that the weights of $\bu$ are:
\begin{center}
\begin{tabular}{|c|c|c|c|c|c|c|c|c|c|c|c|}\hline
                & $u_1$ & $u_2$  &  $u_3$ & $u_4$ & $u_5$ & $u_1$ \\ \hline
\textbf{Weight} & $+11$ & $+7$   &  $+6$  & $+3$  & $+2$  & $+1$  \\ \hline
\end{tabular}
\end{center}
Next we choose a basis of cycles (closed paths) upon the surface defined by $C$.  We denote them 
\begin{equation*}
\alpha_i, \beta_j, \qquad 1 \leq i,j \leq 6,
\end{equation*}
and ensure they have intersection numbers
\begin{equation*}
\alpha_i \cdot \alpha_j = 0, \qquad \beta_i \cdot \beta_j = 0, \qquad 
\alpha_i \cdot \beta_j = \delta_{ij} = \Big\{  
\begin{array}{ccc}
1 & \mbox{if} & i = j \\
0 & \mbox{if} & i \neq j
\end{array}.  
\end{equation*}

Let $\Lambda$ denote the lattice generated by the integrals of the basis of holomorphic differentials around this basis of cycles in $C$.  Then the manifold $\C^6 / \Lambda$ is the Jacobian variety of $C$, denoted by $J$.  Next, for $k=1,2,\dots$ define $\mathfrak{A}$, the \emph{Abel map} from the $k$th symmetric product Sym$^k(C)$ to $J$.
\begin{align}
\mathfrak{A}: \mbox{Sym}^k(C) &\to     J \nonumber \\
(P_1,\dots,P_k)   &\mapsto \left( \int_{\infty}^{P_1} \bm{du} + \dots + \int_{\infty}^{P_k} \bm{du} \right) \pmod{\Lambda},
\label{Abel}
\end{align}
where the $P_i$ are again points upon $C$.  Denote the image of the $k$th Abel map by $W^{[k]}$, and let
$[-1](u_1, \dots ,u_6) = (-u_1, \dots ,-u_6)$.  We then define the \emph{$k$th standard theta subset} (also referred to as the $k$th stratum) by
\begin{equation*}
\Theta^{[k]} = W^{[k]} \cup [-1]W^{[k]}.
\end{equation*}
When $k=1$ the Abel map gives a one dimensional image of the curve $C$.  Since our mapping was given by a single integral with respect to one parameter, it will make sense to rewrite this as an integral on the one-dimensional stratum, $\Theta^{[1]}$.  In addition to \cite{bg03},\cite{bg04} and \cite{bg06}, similar problems of inverting meromorphic differentials on lower dimensional strata of the Jacobian have been studied, in the case of hyperelliptic surfaces, in \cite{DoublePend}, \cite{HL08}, \cite{AbF00} and \cite{AlF00} for example.

\quad

Let us also define a basis of second kind meromorphic differentials, $\bm{dr}$,  for the surface; these have their only pole at $\infty$. These are determined modulo the space spanned by the $\bm{du}$ and can be expressed as
\begin{equation*} 
\bm{dr} = (dr_1,\dots,dr_6), \qquad \mbox{where} \quad dr_j(t,s) = \frac{h_j(t,s)}{4s^3}dx.
\end{equation*}
A specific set was derived in \cite{MYPAPER} in order to construct Klein's explicit realisation of the fundamental differential of the second kind.  This set was given as
\begin{align*}
h_1 &= -s^2 \big(  11t^3 + 8t^2\mu_4 + 5t\mu_3 + 2\mu_2 \big), \qquad
h_2 = -s^2 \big( 7t^2 + 4t\mu_4 +\mu_3 \big), \nonumber \\
h_3 &= -2ts \big(  3t^2 + 2t\mu_4+\mu_3  \big), \quad
h_4 = -3ts^2, \quad
h_5 = -2t^2s, \quad
h_6 = -t^3.
\end{align*}

We then define the period matrices $\w',\w'',\eta'$ and $\eta''$ by
\begin{align*}
\begin{array}{cc}
          2\w'  = \left( \oint_{\alpha_k} du_\ell \right)_{k,\ell = 1,\dots,6} &
\qquad    2\w'' = \left( \oint_{ \beta_k} du_\ell \right)_{k,\ell = 1,\dots,6}  \\
        2\eta'  = \left( \oint_{\alpha_k} dr_{\ell} \right)_{k,\ell = 1,\dots,6} &
\qquad  2\eta'' = \left( \oint_{ \beta_k} dr_{\ell} \right)_{k,\ell = 1,\dots,6}
\end{array}.
\end{align*}
We can combine these into 
\begin{align*}
M =
\left( \begin{array}{cc}
\w'   & \w''        \\
\eta' & \eta''
\end{array} \right) ,
\end{align*}
which satisfies the generalised Legendre relation,
\[
M \left( \begin{array}{cc}
0   & -1_6 \\
1_g & 0
\end{array} \right) M^T = -\frac{i \pi}{2} \left( \begin{array}{cc}
0   & -1_6 \\
1_g & 0
\end{array} \right).
\]

\section{The Kleinian $\sigma$-function associated with $\bm{C}$} \label{SEC_SIG}

We will now define the multivariate $\sigma$-function associated with $C$,
which is constructed from the $\theta$-function, (see for example, \cite{mu83}).
This function is a generalisation of the classical Weierstrass
elliptic $\sigma$-function, and as in the elliptic case, it can be used to
construct Abelian functions on $J$.
\begin{definition} \label{sigdef}
The \emph{Kleinian $\s$-function associated with $C$} is
\begin{align*}
\s(\textbf{u}) &= \sigma(\bu; M) = c \exp \big(\textstyle -\frac{1}{2} \bm{u} \eta' (\w')^{-1} \bm{u}^T \big) \times \theta[\delta]\big((\w')^{-1}\bm{u}^T \hspace*{0.05in} \big| \hspace*{0.05in} (\w')^{-1} \w''\big) \nonumber \\
 &= c \exp \big( - \textstyle \frac{1}{2} \bm{u} \eta' (\w')^{-1} \bm{u}^T \big) 
\times \displaystyle \sum_{m \in \Z^6} \exp \bigg[ \\ 
&\quad 2\pi i \bigg\{ \textstyle \frac{1}{2} (m+\delta')^T (\w')^{-1} \w''(m+\delta') + 
(m+\delta')^T ((\w')^{-1} \bm{u}^T + \delta'') \bigg\}  \bigg]. \nonumber
\end{align*}
Here $c$ is a constant dependent upon the curve parameters, $(\mu_0, \mu_1, \mu_2, \mu_3, \mu_4)$. The results of this paper are independent of this
constant, so we do not discuss its value here.  The matrix 
$\delta = \left[ \begin{array}{l}
\bm{\delta'} \\ \bm{\delta''}
\end{array}
\right]$
is the theta function characteristic which gives the Riemann constant for $C$ with respect to the base point $\infty$ and the period matrix $[\w', \w'']$, (see \cite{bel97} p23-24).
\end{definition}
\noindent We will evaluate the integrand using derivatives of 
$\sigma(\bu)$, with respect to the variables $\bu$.  We denote these 
$\sigma$-derivatives by adding subscripts.  For example we write
$ \frac{\partial \sigma}{\partial u_i}$ as $\sigma_i.$

\begin{lemma} \label{sigprop}
We summarise the fundamental properties of the $\sigma$-function in this
lemma.  Further details and proofs are available in \cite{MYPAPER}.  For
a detailed study of the general multivariate $\sigma$-function, we refer
the reader to \cite{bel97}.
\begin{itemize}
\item Given $\bu \in \C^6$, denote by $\bm{u'}$ and $\bm{u''}$ the unique
elements in $\R^6$ such that 
$\bu = \bm{u'}\w' + \bm{u''}\w''$.  Let $\ell$ represent a point on the period
lattice
\begin{equation*}
\ell = \ell'\w' + \ell''\w'' \in \Lambda.
\end{equation*}
For $\bu, \bv \in \C^6$ and $\ell \in \Lambda$, define $L(\bu,\bv)$ and 
$\chi(\ell)$ as follows:
\begin{align*}
L(\bu,\bv) &= \bu^T \big( \eta'\bm{v'} + \eta''\bm{v''} \big), \\
\chi(\ell) &= \exp \big[ \pi i \big( 2(\ell'^T\delta'' - \ell''^T\delta') + \ell'^T\ell'' \big) \big].
\end{align*}
Then, for all $\bu \in \C^6, \ell \in \Lambda$ the function $\sigma(\bu)$ is quasi-periodic.
\begin{align} \label{quas}
\sigma(\bu + \ell) = \chi(\ell) \exp \Big[ L \Big( \bu + \frac{\ell}{2}, \ell \Big) \Big] \cdot \sigma(\bu).
\end{align}
\item For $\gamma \in Sp(12,\Z)$ we have
\begin{equation} \label{Mper}
\sigma(\bu; \gamma M) = \sigma(\bu; M).
\end{equation}
\item $\sigma(\bu)$ has zeroes of order 1 when $\bu \in \Theta^{[5]}$, and is non-zero elsewhere.
\item In the case when all the curve parameters are set to zero, the function $\sigma(\bu)$ is equal to a constant $\mathcal{K}$ times the Schur-Weierstrass polynomial,
\begin{align}
&SW_{4,5} =  \textstyle \frac{1}{8382528}u_{6}^{15} + \frac{1}{336}u_{6}^{8}u_{5}^{2}u_{4} 
- \frac{1}{12}u_{6}^{4}u_{1} - \frac{1}{126}u_{6}^{7}u_{3}u_{5} - \frac{1}{6}u_{4}u_{3}u_{5}u_{6}^{4} \nonumber \\
&\quad \textstyle - \frac{1}{72}u_{4}^{3}u_{6}^{6} - \frac{1}{33264}u_{6}^{11}u_{5}^{2} 
+ \frac{1}{27}u_{5}^{6}u_{6}^{3} + \frac{2}{3}u_{4}u_{5}^{3}u_{3} - 2u_{4}^{2}u_{6}u_{3}u_{5} - u_{2}^{2}u_{6} \nonumber \\
&\quad \textstyle - \frac{2}{9}u_{5}^{3}u_{3}u_{6}^{3} - u_{4}u_{3}^{2} + \frac{1}{12}u_{4}^{4}u_{6}^{3} 
- \frac{1}{3024}u_{6}^{9}u_{4}^{2} - \frac{1}{756}u_{6}^{7}u_{5}^{4} + \frac{1}{1008}u_{6}^{8}u_{2} \nonumber \\
&\quad \textstyle + \frac{1}{3}u_{5}^{4}u_{2} + \frac{1}{3}u_{6}^{3}u_{3}^{2} - \frac{1}{9}u_{4}u_{5}^{6} 
+ \frac{1}{399168}u_{6}^{12}u_{4} + u_{4}u_{6}u_{5}^{2}u_{2} + \frac{1}{4}u_{4}^{5} \nonumber \\
&\quad \textstyle + 2\,u_{{5}}u_{{3}}u_{{2}} + \frac{1}{6}\,{u_{{5}}}^{2}{u_{{6}}}^{4}u_{{2}} 
+ \frac{1}{12}\,{u_{{6}}}^{5}u_{{2}}u_{{4}} - \frac{1}{2}\,{u_{{4}}}^{2}{u_{{6}}}^{2}u_{{2}} 
+ \frac{1}{2}\,{u_{{4}}}^{3}{u_{{6}}}^{2}{u_{{5}}}^{2} \nonumber \\  
&\quad \textstyle - \frac{1}{3}\,{u_{{4}}}^{2}u_{{6}}{u_{{5}}}^{4}  - \frac{1}{36}\,{u_{{5}}}^{4}u_{{4}}{u_{{6}}}^{4} 
+ u_{{4}}u_{{6}}u_{{1}} - {u_{{5}}}^{2}u_{{1}}. \label{SW45}
\end{align}
\item The sigma function must have definite parity and weight.  From $SW_{45}$ above we can conclude that $\sigma(\bu)$ is odd with weight $+15$.
\end{itemize}
\end{lemma}

One of the key results in \cite{MYPAPER} was a Taylor series expansion for $\sigma(\bu)$ about the origin, used to derive relations between the Abelian functions associated with $C$.  We will use this expansion in the evaluation of the integrand given in Section \ref{SEC_EI}.  In \cite{MYPAPER} it was shown that this expansion could be constructed in the form 
\[
\sigma(\textbf{u}) = \sigma(u_1, u_2, u_3, u_4, u_5, u_6) = C_{15}(\bu) +  C_{19}(\bu) +  \dots + C_{15 + 4n}(\bu) + \dots
\]
where each $C_k$ was a finite, odd isobaric polynomial composed of sums of
monomials in $u_i$ of total weight $+k$, each multiplied by a monomial 
in $\mu_j$ of total weight $15-k$.  From Lemma \ref{sigprop} we can 
conclude that $C_{15}=SW_{4,5}$,
while the other $C_k$ were found in turn, up to $C_{59}$ by considering the
possible terms, and ensuring the expansion satisfied known properties of
$\sigma(\bu)$.  (See \cite{MYPAPER} for full details of the construction,
and \cite{Bweb} for a link to the expansion.)

\section{Deriving relations between the $\sigma$-function and its derivatives} \label{SEC_Sig_Rel}

We will evaluate the integrand (\ref{integrand2}) as a function of $\sigma$-derivatives restricted to $\Theta^{[1]}$.  In order to achieve this we will need to derive equations that hold between the various $\sigma$-derivatives.  In \cite{MYPAPER}, sets of relations between the Abelian functions associated with $C$ were calculated.  However, these were the relations that held everywhere on $J$, and do not give us sufficient information for the behaviour of $\sigma(\bu)$ on the strata.  To derive such relations we start by considering a theorem of 
Jorgenson \cite{jorg92}:

\begin{theorem} \label{Jorgenson} 
Let $\bu \in \Theta^{[k]}$ for some $k<g$.  Then for a set of $k$ points $P_i=(t_i,s_i)$ on $C$ we have
\[
\bu = \sum_{i=1}^k \int_{\infty}^{P_i} \bm{du},
\]
and the following statement holds for vectors $\bm{a},\bm{b}$ of arbitrary constants.
\[
\frac{ \sum_{j=1}^g a_j \sigma_j(\bu) }{ \sum_{j=1}^g b_j \sigma_j(\bu) }
= \frac{\det\big[\bm{a} \big| \bm{du}(P_1) \big| \cdots \big| \bm{du}(P_k) \big| 
\bm{du}(P_k)^{(g-k-1)} \big| \cdots \big| \bm{du}(P_k)^{(1)} \big] } 
{\det\big[\bm{b} \big| \bm{du}(P_1) \big| \cdots \big| \bm{du}(P_k) \big| 
\bm{du}(P_k)^{(g-k-1)} \big| \cdots \big| \bm{du}(P_k)^{(1)} \big] }
\]
Here, $\bm{du}^{(i)}$ denotes the column of $i$th derivatives of the holomorphic differentials $\bm{du}$, and should be ignored if $i<1$.
\end{theorem}
\noindent Below we state that the strata of $C$ can be defined by the zeros
of the $\sigma$-function, and of its derivatives.  The definition of $\Theta^{[5]}$
is a classical result (in Lemma \ref{sigprop}) while the others can be derived
from the theorem above.  (See Appendix \ref{App_Jorg} for full details.)
\begin{align}  
\Theta^{[5]} &= \{ \bu \hspace*{0.05in} | \hspace*{0.05in} \sigma(\bu) 
= 0 \} \nonumber \\
\Theta^{[4]} &= \{ \bu \hspace*{0.05in} | \hspace*{0.05in} \sigma(\bu) =
\sigma_6(\bu) = 0 \}  \nonumber \\
\Theta^{[3]} &= \{ \bu \hspace*{0.05in} | \hspace*{0.05in} \sigma(\bu) =
 \sigma_6(\bu) = \sigma_5(\bu) = 0 \} \label{Strata_Def_Rel} \\
\Theta^{[2]} &= \{ \bu \hspace*{0.05in} | \hspace*{0.05in} \sigma(\bu) 
= \sigma_6(\bu) = \sigma_5(\bu) = \sigma_4(\bu) = 0 \}  \nonumber \\
\Theta^{[1]} &= \{ \bu \hspace*{0.05in} | \hspace*{0.05in} \sigma(\bu) 
= \sigma_6(\bu) = \sigma_5(\bu) = \sigma_4(\bu) = \sigma_3(\bu) = 0 \} \nonumber
\end{align}
We use these defining relations to generate further relations between the
$\sigma$-derivatives, holding on each stratum.  We use a systematic method,
implemented in Maple, to achieve this.

Start with the relation $\sigma(\bu)=0$ valid for $\bu \in \Theta^{[5]}$.
Consider $\bu$ as it descends to $\Theta^{[4]}$.  We write $\bu$ as 
$\bu = \bm{\hat{u}} + \bm{u_{\xi}}$ where $\bm{\hat{u}}$ is an arbitrary
point on $\Theta^{[4]}$ and $\bm{u_{\xi}}$ is a vector containing the series
expansions, (\ref{uxi}). We can calculate the Taylor series expansion in
$\xi$ for $\sigma(\bm{\hat{u}} + \bm{u_{\xi}})=0$ as
\begin{align*}
0 &= \sigma(\bm{\hat{u}} + \bm{u_{\xi}}) = 
\textstyle \sigma(\bm{\hat{u}}) - \sigma_{6}(\bm{\hat{u}})\xi 
+ \frac{1}{2}\big[ \sigma_{66}(\bm{\hat{u}}) 
- \sigma_{5}(\bm{\hat{u}}) \big] \xi^2 
+ \big[ \frac{1}{2}\sigma_{56}(\bm{\hat{u}}) \\
&\quad \textstyle - \frac{1}{3}\sigma_{4}(\bm{\hat{u}}) 
- \frac{1}{6}\sigma_{666}(\bm{\hat{u}}) \big]{\xi}^{3} 
+ \big[ \frac{1}{8}\sigma_{55}(\bm{\hat{u}}) 
+ \frac{1}{3}\sigma_{46}(\bm{\hat{u}}) 
- \frac{1}{4}\sigma_{566}(\bm{\hat{u}}) \\
&\quad \textstyle + \frac{1}{24}\sigma_{6666}(\bm{\hat{u}}) \big]\xi^4 
+ \big[ \frac{1}{12}\sigma_{5666}(\bm{\hat{u}})
+ \frac{1}{6}\sigma_{45}(\bm{\hat{u}}) 
- \frac{1}{120}\sigma_{66666}(\bm{\hat{u}}) 
- \frac{1}{6}\sigma_{466}(\bm{\hat{u}}) \\
&\quad \textstyle - \frac{1}{8}\sigma_{556}(\bm{\hat{u}}) 
+ \frac{1}{20}\sigma_{6}(\bm{\hat{u}})\mu_{4} \big] {\xi}^{5} + O(\xi^6)
\end{align*}
Setting the coefficients of $\xi$ to zero gives us a set of relations for $\bu \in \Theta^{[4]}$:
\begin{align}
\sigma_{6}(\bu) &= 0 \nonumber \\ 
\sigma_{66}(\bu) &= \sigma_{5}(\bu) \nonumber\\
\sigma_{666}(\bu) &= 3\sigma_{56}(\bu) 
- 2\sigma_{4}(\bu) \label{RelThet4} \\
\sigma_{6666}(\bu) &= 6\sigma_{566}(\bu) - 8\sigma_{46}(\bu) 
- 3\sigma_{55}(\bu) \nonumber \\
\sigma_{66666}(\bu) &= 10\sigma_{5666}(\bu) - 20\sigma_{466}(\bu) 
- 15\sigma_{556}(\bu) + 20\sigma_{45}(\bu) 
+ 6\mu_4\sigma_{6}(\bu) \nonumber \\
&\hspace*{0.07in} \vdots \nonumber 
\end{align}
If we calculate the expansion to a higher order of $\xi$ then more relations
can be obtained.  Note however, that the expansion for $\bm{u_{\xi}}$ must
first be calculated to a sufficiently high order first.  
We have calculated an
expansion for $\sigma(\bm{\hat{u}} + \bm{u_{\xi}})$ up to $O(\xi^{29})$,
using the weight properties of $\sigma(\bu)$ to simplify the calculation.
This expansion can be found on-line at \cite{Bweb}.

The next step in this process will be to find the relations valid for 
$\bu \in \Theta^{[3]}$.  Since 
$\Theta^{[3]} \subset \Theta^{[4]}$ we can conclude that the relations 
(\ref{RelThet4}) are valid here also.  However, we can derive a larger set
of relations for $\bu \in \Theta^{[3]}$ by repeating the descent procedure
 for those relations that are valid on $\Theta^{[4]}$.

We do not need to consider the relation $\sigma(\bu)=0$ since that will only
give us the same relations as above.  Instead choose the second defining
relation $\sigma_6(\bu)=0$.  We again write $\bu = \bm{\hat{u}} + \bm{u_{\xi}}$
where $\bm{u_{\xi}}$ is the vector of expansions as before and $\bm{\hat{u}}$
is now an arbitrary point on $\Theta^{[3]}$.  We do not need to calculate
the Taylor series expansion in $\xi$ for $\sigma_6(\bm{\hat{u}} + \bm{u_{\xi}})$
as before.  Instead we can take the previous expansion and simply add 6 to
each index:
\begin{align*}
0 &= \textstyle \sigma_{6}(\bm{\hat{u}}) - \sigma_{66}(\bm{\hat{u}})\xi 
+ \frac{1}{2}\big[ \sigma_{666}(\bm{\hat{u}}) - \sigma_{56}(\bm{\hat{u}}) \big] \xi^2  
+ \big[ \frac{1}{2}\sigma_{566}(\bm{\hat{u}}) - \frac{1}{3}\sigma_{46}(\bm{\hat{u}})  \\
&\quad \textstyle - \frac{1}{6}\sigma_{6666}(\bm{\hat{u}}) \big]{\xi}^{3} 
+ \dots
\end{align*}
Setting the coefficients of $\xi$ to zero gives us  more relations valid for $\bu \in \Theta^{[3]}$, starting with
$\sigma_{66}(\bu) = 0$. We can obtain further relations for $\bu \in \Theta^{[3]}$ by descending all of (\ref{RelThet4}). We automate this process in Maple as follows:
\begin{enumerate}[1.]
\item Take a relation valid for $\bu \in \Theta^{[4]}$ and expand as a Taylor series in $\xi$.  To do this we replace each $\sigma$-derivative by the Taylor series expansion for $\sigma(\bm{\hat{u}} + \bm{u_{\xi}})$, adding the relevent index to each $\sigma$-derivative in the expansion.
\item Set each coefficient with respect to $\xi$ to zero, and save the resulting equations.  
\item Repeat steps 1 and 2 for all known relations valid for $\bu \in \Theta^{[4]}$.
\item Use the set of equations we have obtained, to express the higher 
index $\sigma$-derivatives using lower-index derivatives.  If we have 
$\sigma$-derivatives with the same number of indices, solve for those 
with the higher indices first.
\end{enumerate}
Once we have finished this process we will have a set of relations valid
for $\bu \in \Theta^{[3]}$.  We can repeat the process by descending each
of these to $\Theta^{[2]}$ creating another set of relations which we can
finally descend to $\Theta^{[1]}$.  We end up with a set of relations valid
for $\bu \in \Theta^{[1]}$, some of which are contained in Appendix \ref{App_RelThet1},
with the full set we have derived available online at \cite{Bweb}.  

The surprising result of these calculations was that on $\Theta^{[1]}$, 
we have $\sigma_{1}(\bu) = \sigma_{2}(\bu) = 0$ along with the other first
derivatives of $\sigma(\bu)$, concluded to be zero using Theorem \ref{Jorgenson}.
These calculations were computationally much more difficult that in \cite{bg06}.
The latter stages were performed in parallel on a small cluster of machines,
using Distributed Maple, (see \cite{DM} and \cite{smb03}).

\section{Evaluating the integrand} \label{SEC_EI}

Recall our integrand, (\ref{integrand2})
\[
\varphi(t) dt = K \left( A_4t^2 + A_3t + A_2 + \frac{A_1}{t} + 
\frac{1}{t^2}  \right) \left( \frac{dt}{4s^3} \right).
\]
Now, $\lambda(t)$ was given by a single integral with respect to one 
parameter, the point $(t,s)$ on $C$. So we rewrite this 
as an integral on the one-dimensional stratum $\Theta^{[1]}$ of $J$, which
we will parametrise by $u_1$.  
We will then evaluate it using $\sigma$-derivatives restricted to 
$\Theta^{[1]}$.   In \cite{bg06} Jorgenson's Theorem was used to 
express $t$ in terms  of $\sigma$-derivatives.  However, if we solved 
(\ref{Jorgenson_t_1}) na\"ively for $\bu \in \Theta^{[1]}$ we would find
\[
t = - \frac{\sigma_1(\bu)}{\sigma_2(\bu)}
\]
which makes no sense given that $\sigma_1(\bu)=0$ for 
$\bu \in \Theta^{[1]}$.  Instead let us take equation (\ref{Jorgenson_t_2})
which was also derived from Jorgenson's Theorem (in Appendix \ref{App_Jorg})
and which holds for $\bu$ on $\Theta^{[2]}$.  We consider what happens to
this as $\bu$ descends to $\Theta^{[1]}$.  We replace $(t_2,s_2)$ with 
the expansions in the parameter $\xi$
and replace the $\sigma$-derivatives by their Taylor series in $\xi$.  If
we then take series expansion of this in $\xi$ and set $\xi=0$ we find that
for $\bu \in \Theta^{[1]}$ we have
\begin{align*}
{\frac {a_{{1}}\sigma_{{23}}(\bu) + 
a_{{2}}\sigma_{{34}}(\bu) }{b_{{1}}\sigma_{{23}}(\bu) 
+ b_{{2}}\sigma_{{34}}(\bu) }} = \frac{a_1t - a_2}{b_1t - b_2}.
\end{align*}
Solving this for $t$ gives:
\begin{equation} \label{t_second}
t = - \frac{\sigma_{23}(\bu)}{\sigma_{34}(\bu)}
\end{equation}
for $\bu \in \Theta^{[1]}$.  Therefore, using the basis of 
differentials (\ref{holodiff}) and equation (\ref{t_second}) we can 
rewrite our integrand as: 
\begin{align}
\varphi(t) dt &= K [ \varphi_1(t) dt + \varphi_2(t) dt ]  \label{varphit}
\end{align}
where
\begin{align}
\varphi_1(t)dt &=  A_2du_1 + A_3du_2 + A_4du_4,  \label{varphi1} \\
\varphi_2(t)dt &=  
\left( \left( \frac{\sigma_{34}(\bu)}{\sigma_{23}(\bu)} \right)^2 
- A_1 \frac{\sigma_{34}(\bu)}{\sigma_{23}(\bu)} \right) du_1  \equiv 
\varphi_2(\bu) du_1. \label{varphi2}
\end{align}
Thus $\varphi_1$ is a sum of holomorphic differentials on $C$, and  $\varphi_2$
is a second kind meromorphic differential. As in 
the previous cases we will need to find a suitable function $\Psi(\bu)$ 
such that
\begin{equation} \label{goal}
\frac{d}{du_1} \Psi(\bu) = \varphi_2(\bu), \qquad \bu \in \Theta^{[1]}.
\end{equation}

We will identify such a function $\Psi(\bu)$ as follows.  First we must derive
the expansions for $\varphi_2(\bu)$ at its poles.  We will then find a function
$\Psi(\bu)$, which has simple poles at the same points as the double 
poles of  $\varphi_2(\bu)$, and which
varies by at worst an additive constant as $\bu$ moves round the $\alpha$ and
$\beta$-cycles of $\Theta_1$.  The function will be chosen so that 
$\frac{d}{du_1}\Psi(\bu)$,  has the same expansion 
at the poles as  $\varphi_2$ and is regular elsewhere.  It then 
follows that the difference $\frac{d}{du_1}\Psi(\bu)-\varphi_2(\bu)$  
is holomorphic and  Abelian; by Liouville's theorem we conclude that 
this difference is a constant.  This constant may be evaluated at any convenient
point.

\subsection{The expansion of $\varphi_2(\bu)$ at the poles} \label{SEC_phi2_u0}

\noindent Recall that $\sigma(\bu)$ was an entire function, and so $\varphi_2(\bu)$ will have poles only when $\sigma_{23}(\bu)=0$.  Since we are working with $\bu \in \Theta^{[1]}$, by (\ref{t_second}) this will occur at the points, one on each sheet, where $t=0$. The cyclic symmetry $[\mbox{i}]$, relating the different
sheets of the curve, acts on $(t,s)$ by $[\mbox{i}](t,s) \mapsto (t,\mbox{i}s)$; hence, it will act on $\bu$ as follows:
\begin{align} \label{sheetcycle}
\begin{array}{lll}
u_1 \mapsto \mbox{i} u_1 & u_2 \mapsto \mbox{i} u_2& u_3 \mapsto - u_3\\
u_4 \mapsto \mbox{i} u_4 & u_5 \mapsto - u_5& u_6 \mapsto -\mbox{i} u_6.
\end{array}
\end{align} 
Let $\bm{u_0}$ be the Abel image of the point on the principal sheet where $\sigma_{23}=0$.  This is the point where  $t=0$ and $s=(\mu_0)^{1/4}$.  Then the full set of zeros of $\sigma_{23}$ are given by $\bm{u_{0,N}} = [\mbox{i}]^N \bm{u_0}$, $N=0,1,2,3$.  We will require the poles to match at all four of these points.

We need to find an expansion for $\varphi_2(\bu)$ at these points.  To start, 
we consider $\bu \in \Theta^{[1]}$ and calculate the Taylor series of 
$\sigma(\bu)$ around the point $\bu = \bm{u_0} = (u_{0,1}, u_{0,2}, u_{0,3}, u_{0,4}, u_{0,5}, u_{0,6})$.  
Writing $w_i = (u_i - u_{0,i})$ we have
\begin{align*}
&\sigma(\bu) = \sigma(\bm{u_0}) + \big[ \sigma_{{1}}(\bm{u_0})w_{{1}} 
+ \sigma_{{2}}(\bm{u_0})w_{{2}}+ \sigma_{{3}}w_{{3}}(\bm{u_0}) 
+ \sigma_{{4}}(\bm{u_0})w_{{4}}  \\
&\textstyle + \sigma_{{5}}(\bm{u_0})w_{{5}} +\sigma_{{6}}(\bm{u_0})w_{{6}} \big]
+ \big[ \frac{1}{2}\sigma_{11}(\bm{u_0})w_1^2 
+ \sigma_{12}(\bm{u_0})w_1w_2 + \sigma_{13}(\bm{u_0})w_1w_3 \\
&\textstyle + \sigma_{14}(\bm{u_0})w_1w_4 
+ \sigma_{15}(\bm{u_0})w_1w_5 + \sigma_{16}(\bm{u_0})w_1w_6
+ \frac{1}{2}\sigma_{22}(\bm{u_0})w_2^2 + \dots 
\end{align*}
We will have similar expansions around the other $\bm{u_{0,N}}$, and we can also use this expansion to easily compute the expansions for the $\sigma$-derivatives, (by simply adding the relevant indices).  Note, that since $\bm{u_{0,N}}$ are the points where $t=0$, we can write their components as
\[
u_{0,i}^{[N]} = \int_{\infty}^0 du_i, \qquad i=1,\dots,6.
\]
evaluated on the sheet where $s=[i]^N(\mu_0)^{1/4}$.  Therefore 
\begin{align*}
w_{i,N} := \big(u_i - u_{0,i}^{[N]}\big) = \int_{\infty}^t du_i - \int_{\infty}^0 du_i = \int_{0}^t du_i,
\end{align*}
evaluated on this sheet.  Using (\ref{holodiff}), our basis of holomorphic differentials, we can find expansions for $w_{1,N},\dots w_{6,N}$ in the parameter $t$. 
\begin{align}
w_{1,N} &= \frac{1}{4}\frac {{{\it i}}^{N}}{{\mu_{{0}}}^{3/4}}t 
- {\frac {3}{32}}{\frac {{{\it i}}^{3N}\mu_{{1}}}{{\mu_{{0}}}^{7/4}}}{t}^{2}
- \frac {{{\it i}}^{3N}}{128}\frac{8\mu_{2}\mu_{0} - 7\mu_{1}^{2}}{{\mu_{0}}^{11/4}}{t}^{3} 
\nonumber \\
&\qquad -\frac{{{\it i}}^{N}}{2048} \frac{(96\mu_{3}{\mu_{0}}^{2} - 168\mu_{1}\mu_{2}\mu_{0} 
+ 77{\mu_{1}}^{3} )} { {\mu_{0}}^{15/4}}{t}^{4} + O \left( {t}^{5} \right)  \label{w1} \\
w_{2,N} &= \frac{1}{8}{\frac {{{\it i}}^{N}}{{\mu_{{0}}}^{3/4}}}{t}^{2} 
- \frac{1}{16}{\frac {{{\it i}}^{N}\mu_{{1}}}{{\mu_{{0}}}^{7/4}}}{t}^{3} 
- \frac {3{{\it i}}^{N}}{512}{\frac {8\mu_{2}\mu_{0} - 7{\mu_{1}}^{2}}{{\mu_{0}}^{11/4}}}{t}^{4}
+ O \left( {t}^{5} \right)  \nonumber \\
w_{3,N} &= \frac{1}{4}{\frac {{{\it i}}^{2N}}{ \mu_{{0}}^{1/2} }}t 
- \frac{1}{16}{\frac {{{\it i}}^{2N}\mu_{{1}}}{{\mu_{{0}}}^{3/2}}}{t}^{2} 
- \frac{{{\it i}}^{2N}}{96}{\frac{4\mu_{{2}}\mu_{{0}} - 3{\mu_{{1}}}^{2}}{{\mu_{{0}}}^{5/2}}}{t}^{3} 
\nonumber \\
& \qquad - {\frac {{{\it i}}^{2N}}{256}}{\frac {8\mu_{{3}}{\mu_{{0}}}^{2} - 12\mu_{{1}}\mu_{{2}}\mu_{{0}} 
+ 5{\mu_{{1}}}^{3}}{{\mu_{{0}}}^{7/2}}}{t}^{4} + O \left( {t}^{5} \right) \nonumber \\
w_{4,N} &= \frac{1}{12}{\frac {{{\it i}}^{N}}{{\mu_{{0}}}^{3/4}}}{t}^{3} 
- {\frac {3}{64}}{\frac {{{\it i}}^{N}\mu_{{1}}}{{\mu_{{0}}}^{7/4}}}{t}^{4} + O \left( {t}^{5} \right) 
\nonumber \\
\end{align}
\begin{align}
w_{5,N} &= \frac{1}{8}{\frac {{{\it i}}^{2N}}{  \mu_{{0}}^{1/2} }}{t}^{2} 
- \frac{1}{24}{\frac {{{\it i}}^{2N}\mu_{{1}}}{{\mu_{{0}}}^{3/2}}}{t}^{3} 
- \frac{{{\it i}}^{2N}}{128}{\frac{4\mu_{{2}}\mu_{{0}} - 3{\mu_{{1}}}^{2}}{{\mu_{{0}}}^{5/2}}}{t}^{4}
+ O \left( {t}^{5} \right)  \nonumber \\ 
w_{6,N} &= \frac{1}{4}{\frac {{\it i}^{3N}}{  \mu_0^{1/4}  }}t 
- \frac{1}{32}{\frac {{\it i}^{3N}\mu_{{1}}}{{\mu_{{0}}}^{5/4}}}{t}^{2} 
- \frac{{\it i}^{3N}}{384}\frac{8\mu_{{2}}\mu_{{0}} - 5{\mu_{{1}}}^{2}}{{\mu_{{0}}}^{9/4}}{t}^{3} 
\nonumber \\
&\qquad - \frac{{\it i}^{3N}}{2048} \frac{(32\mu_{{3}}{\mu_{{0}}}^{2} 
- 40\mu_{{1}}\mu_{{2}}\mu_{{0}} + 15{\mu_{{1}}}^{3})} { \mu_{0}^{ 13/4 }}{t}^{4} 
+ O \left( {t}^{5} \right)  \nonumber
\end{align}
Note that all these expansions are given for the general sheet since we need to check the behaviour at all the poles.  We can move between the sheets by selecting the appropriate value of $N$.  We can invert (\ref{w1})  on the $N$-th sheet to give an expansion for $t$ in $w_{1,N}$, allowing us to use $w_{1,N}$ as a local parameter near $\bm{u_{0,N}}$.
\[
t = 4\,{{\it i}}^{3N}{\mu_{{0}}}^{3/4}w_{1,N} + 6\mu_{{1}}{{\it i}}^{6N} \mu_{{0}}^{1/2}  w_{1,N}^{2}
+O \left( w_{1,N}^{3} \right). 
\]
We start by substituting for $t$ to give the expansions of $w_{2,N},\dots,w_{6,N}$ with respect to $w_{1,N}$.
\begin{align}
w_{2,N} &= 2{{\it i}}^{3N}{\mu_{{0}}}^{3/4}w_{1,N}^{2} + O \left( w_{1,N}^{3} \right) \nonumber \\
& \vdots \label{wi_w1} \\
w_{6,N} &= {{\it i}}^{2N} \mu_{{0}}^{1/2} w_{1,N} + \mu_{{0}}^{1/4}\mu_{{1}}{{\it i}}^{5N} w_{1,N}^{2}
+ O \left( w_{1,N}^{3}  \right). \nonumber
\end{align}
We use these in turn to give the $\sigma$-derivative expansions at $\bm{u_{0,N}}$ as series in $w_{1,N}$. For example we have, 
\begin{align*}
\sigma_{23}(\bu) &= \sigma_{23}(\bm{u_{0,N}}) 
+ \left( {{\it i}}^{2N}\mu_{0}^{\frac{1}{2}}\sigma_{{236}} + \sigma_{{123}} 
+ {{\it i}}^{N}\mu_{{0}}^{\frac{1}{4}}\sigma_{{233}} \right)(\bm{u_{0,N}})  w_{1,N} \\
&\quad + O \left( w_{1,N}^{2} \right).
\end{align*}
We substitute these into (\ref{varphi2}) to obtain an expansion of $\varphi_2(\bu)$ at $\bu = \bm{u_{0,N}}$, as a series in $w_{1,N}$. 
% \[
% \frac{\sigma_{34}}{\sigma_{23}} = \frac{ \sigma_{34} }{ \sigma_{236}\sqrt{\mu_{{0}}}{{\it i}}^{2\,N}+\sigma_{{123}} +\sigma_{{233}}\sqrt [4]{\mu_{{0}}}{{\it Ii}}^{N} }  \left( \frac{1}{w_1} \right) + O\left( 1 \right) 
% \]
\begin{align*}
\varphi_2(\bu) &= \frac{1}{w_{1,N}^2}\left( \frac{ \sigma_{34} }{ {\it i^N}\mu_{{0}}^{1/4}\sigma_{{233}}
+ {{\it i}}^{2N} \mu_{{0}}^{1/2} \sigma_{{236}} + \sigma_{{123}}  }\right)^2 (\bm{u_{0,N}})   \\
&\qquad +  \frac{ \mathcal{C}(\bm{u_{0,N}}) }{w_{1,N}} + O(w_{1,N}^0)
\end{align*}
where $\mathcal{C}(\bm{u_{0,N}})$ is a polynomial in the $\sigma$-derivatives, which we need to evaluate to ensure that $\varphi_2 dt$ has zero residue.

In the previous section we derived a set of relations for $\bu \in \Theta^{[1]}$,
but these are not sufficient to simplify $\mathcal{C}(\bm{u_{0,N}})$.  We need to generate
a further set of relations which are valid only at $\bu = \bm{u_{0,N}}$.
We do this using a similar approach to the previous section.
We take a relation valid on $\Theta^{[1]}$ and calculate its expansion
around $\bm{u} = \bm{u_{0,N}}$ as a series in $w_{1,N}$, using the series derived
above.  We then set to zero the coefficients of $w_{1,N}$.  
We do this for each relation valid on $\Theta^{[1]}$ and obtain a set 
of equations between $\sigma$-derivatives at the points $\bu = \bm{u_{0,N}}$.
The first few such relations are given in Appendix \ref{APP_Rel_u0}, with
a fuller set available online at \cite{Bweb}.

If we substitute these into the expansion of $\varphi_2(\bu)$ we obtain
\begin{align}
\varphi_2(\bu) = \left[ \frac{{\it i}^{2N}}{16} \frac{1}{\mu_0^{3/2}} \right]\frac{1}{w_{1,N}^2} 
+ \left[ \frac{{\it i}^N}{16}  (4\mu_0A_1 - 3\mu_1)\right]\frac{1}{w_{1,N}} + O(w_{1,N}^0) \label{varphi2_u0}
\end{align}
Recall equation (\ref{zr1}) which stated that $\varphi(p)$ has zero 
residue at $p=\infty$ on all sheets.  Since residues are invariant 
under conformal maps, we can conclude that $\varphi_2(\bu)$ must also 
have zero residue, and so the constant $A_1$ must be equal to
\begin{equation}
A_1 = \frac{3}{4}\frac{\mu_{1}}{\mu_0}, 
\qquad \mbox{giving} \qquad  \label{varphi2_pole}
\varphi_2(\bu) = 
\left[ \frac{{\it i}^{2N}}{16}  
\frac{1}{\mu_0^{3/2}} \right]\frac{1}{w_{1,N}^2}  + O(w_{1,N}^0).
\end{equation}

\subsection{Finding a suitable function $\Psi(\bu)$} \label{SEC_Psi}

We need to derive a function $\Psi(\bu)$ such that the Laurent expansion
of $\frac{\mathrm{d}\Psi(\bu)}{\mathrm{d}u_1}$ has the same principal
part at the poles as $\varphi_2(\bu)$,  so we will restrict our search to linear
expressions in  $\sigma$-derivatives, divided by $\sigma_{23}(\bu)$.  
For these functions we will derive expansions in $w_{1,N}$ at 
$\bu = \bm{u_{0,N}}$ using the techniques described above.  Let us take the function
\[
\Psi(\bu) =  
\sum_{1 \leq i \leq 6} \eta_{i}\frac{\sigma_{i}(\bu)}{\sigma_{23}(\bu)}
+ \sum_{  \substack{1 \leq i \leq j \leq 6 \cr [i,j] \neq [2,3]}  } \eta_{ij}\frac{\sigma_{ij}(\bu)}{\sigma_{23}(\bu)} 
+ \sum_{1 \leq i \leq j \leq k \leq 6} \eta_{ijk}\frac{\sigma_{ijk}(\bu)}{\sigma_{23}(\bu)},
\]
where the $\eta_{ij}$ and $\eta_{ijk}$ are undetermined constants.   
(Through trial and error we found that 3-index $\sigma$-derivatives were
required in the numerator).  Now, since we are working on $\Theta^{[1]}$
we will find that many of these $\sigma$-derivatives are equal to zero, or
can be expressed as a linear combination of other such functions using the
equations in Appendix \ref{App_RelThet1}.  Let us set the coefficients of
these functions to zero, leaving us with 
\begin{align*}
&\Psi(\bu) = [ \eta_{22}\sigma_{22} + \eta_{34}\sigma_{34} 
+ \eta_{111}\sigma_{111} + \eta_{122}\sigma_{122} 
+ \eta_{123}\sigma_{123} + \eta_{134}\sigma_{134}   \\ 
&\quad + \eta_{222}\sigma_{222} + \eta_{223}\sigma_{223} + \eta_{224}\sigma_{224} 
+ \eta_{225}\sigma_{225} + \eta_{226}\sigma_{226} 
+ \eta_{233}\sigma_{233}  \\
&\quad + \eta_{234}\sigma_{234} 
+ \eta_{235}\sigma_{235} + \eta_{236}\sigma_{236} + \eta_{334}\sigma_{334} 
+ \eta_{344}\sigma_{344} + \eta_{345}\sigma_{345} \\
&\quad + \eta_{346}\sigma_{346}
](\bu) \cdot \frac{1}{\sigma_{23}(\bu)}.
\end{align*}
We emphasise that we need to work with the {\em total}, not the 
partial, derivative of $\Psi(\bu)$ with respect to $u_1$; in practice 
the other $u_i$ are expressed in terms of $w_{1,N}$ in the vicinity of $\bm{u_{0,N}}$ so there is no ambiguity.
Note from (\ref{holodiff}) that
\begin{align*}
% \begin{array}{lll}
% \displaystyle \frac{\partial}{\partial u_2} = t \frac{\partial}{\partial u_1},  &
% \frac{\partial}{\partial u_3} = s \frac{\partial}{\partial u_1},  &
% \frac{\partial}{\partial u_4} = t^2 \frac{\partial}{\partial u_1}, \\
% \frac{\partial}{\partial u_5} = st \frac{\partial}{\partial u_1},  &
% \frac{\partial}{\partial u_6} = s^2 \frac{\partial}{\partial u_1}. &
% \end{array}
\textstyle \frac{\partial}{\partial u_2} = t \frac{\partial}{\partial u_1},   \quad
\textstyle \frac{\partial}{\partial u_3} = s \frac{\partial}{\partial u_1},   \quad
\textstyle \frac{\partial}{\partial u_4} = t^2 \frac{\partial}{\partial u_1}, \quad
\textstyle \frac{\partial}{\partial u_5} = st \frac{\partial}{\partial u_1},  \quad
\textstyle \frac{\partial}{\partial u_6} = s^2 \frac{\partial}{\partial u_1}.
\end{align*}
Therefore
\begin{align*}
D_1 &:= \frac{d}{du_1} \Big|_{\Theta^{[1]}} = \frac{\partial}{\partial u_1}
+ t\frac{\partial}{\partial u_2} 
+ s\frac{\partial}{\partial u_3} 
+ t^2\frac{\partial}{\partial u_4} + st\frac{\partial}{\partial u_5}
+ s^2 \frac{\partial}{\partial u_6} \\
&= \frac{\partial}{\partial u_1} 
- \frac{\sigma_{23}(\bu)}{\sigma_{34}(\bu)}\frac{\partial}{\partial u_2} 
+ s\frac{\partial}{\partial u_3} 
+ \left(\frac{\sigma_{23}(\bu)}{\sigma_{34}(\bu)}\right)^2
\frac{\partial}{\partial u_4} 
- s\frac{\sigma_{23}}{\sigma_{34}}\frac{\partial}{\partial u_5}
+ s^2 \frac{\partial}{\partial u_6} 
\end{align*}
We can now evaluate $\frac{d}{du_1} \Psi(\bu) $ as a sum of quotients 
of $\sigma$-derivatives.  For example 
\begin{align*}
&D_1 \bigg( \frac{\sigma_{236}}{\sigma_{23}} \bigg) 
= \bigg[ \frac{\sigma_{1236}}{\sigma_{23}} 
- \frac{\sigma_{236}\sigma_{123}}{\sigma_{23}^2} 
- \frac{\sigma_{2236}}{\sigma_{34}} + \frac{\sigma_{236}\sigma_{223}}{\sigma_{23}\sigma_{34}}
+ s \frac{\sigma_{2336}}{\sigma_{23}} 
- s\frac{\sigma_{236}\sigma_{233}}{\sigma_{23}^2}   \\ 
&\quad
+ \frac{\sigma_{23}\sigma_{2346}}{\sigma_{34}^2} 
- \frac{\sigma_{236}\sigma_{234}}{\sigma_{34}^2}   
- s\frac{\sigma_{2356}}{\sigma_{34}^2} 
+ s\frac{\sigma_{236}\sigma_{235}}{\sigma_{23}\sigma_{34}}  
+ s^2 \frac{\sigma_{2366}}{\sigma_{23}} 
- s^2\frac{\sigma_{236}^2}{\sigma_{23}^2} \bigg].
\end{align*}

Now let us consider the expansion of $D_1(\Psi(\bu))$ at $\bu = \bm{u_{0,N}}$.
We generate series expansions in $w_{1,N}$ for the relevant $\sigma$-derivatives
using the method described in the previous subsection.  We can use the relations
in Appendix \ref{App_RelThet1} and \ref{APP_Rel_u0} to simplify these expansions, and so obtain a series
in $w_{1,N}$ for $D_1(\Psi(\bu))$.  We find,
\[
\frac{d}{du_1} \Psi(\bu) \quad \Big|_{\bu = \bm{u_{0,N}}} = 
\frac{N(\bm{u_{0,N}})}{\sigma_{22}(\bm{u_{0,N}})} 
\left[ \frac{1}{w_{1,N}^2} \right] + O(w_{1,N}^0),
\]
where $N(\bm{u_{0,N}})$ is a linear polynomial in 
$\{ \sigma_{22}, \sigma_{122}, \sigma_{222}, \sigma_{223}, \sigma_{224},
\sigma_{225}, \sigma_{226} \}$. This set of $\sigma$-derivatives can be used
to express all other 2 and 3-index $\sigma$-derivatives when $\bu = \bm{u_{0,N}}$,
(as in Appendix \ref{APP_Rel_u0}).  We find the coefficients of $N(\bm{u_{0,N}})$
with respect to each of these seven $\sigma$-derivatives and determine conditions
on the constants $\eta_{ij},\eta_{ijk}$ that set all the coefficients, except
that of $\sigma_{22}$, to zero.  We then obtain further conditions by ensuring
the expansion we are left with (now independent of any $\sigma$-derivatives)
is equal to (\ref{varphi2_u0}) on the four sheets.  Imposing these conditions
on $\Psi(\bu)$ leaves us with
\[
\Psi(\bu) 
= \bigg[ \eta_{22}\frac{\sigma_{22}}{\sigma_{23}} 
+ \eta_{111}\frac{\sigma_{111}}{\sigma_{23}} 
+ 2\eta_{334}\frac{\sigma_{235}}{\sigma_{23}} 
+ \frac{8\eta_{22}\mu_0 - 1}{4\mu_0}
\frac{\sigma_{236}}{\sigma_{23}} 
+ \eta_{334}\frac{\sigma_{334}}{\sigma_{23}} \bigg](\bu).
\]
Note from Appendix \ref{APP_Rel_u0} that the terms containing 
$\sigma_{111}, \sigma_{235}, \sigma_{334}$ all vanish at the points 
$\bu = \bm{u_{0,N}}$ and so have no effect on the expansion here.  
Let us discard these to leave,
\[
\Psi(\bu) = \eta_{22}\frac{\sigma_{22}(\bu)}{\sigma_{23}(\bu)} 
+ \frac{1}{4}\frac{(8\eta_{22}\mu_0 - 1)}{\mu_0}\frac{\sigma_{236}(\bu)}{\sigma_{23}(\bu)}
\]
We now have two functions, $\varphi_2(\bu)$ and $D_1(\Psi(\bu))$, 
which both have poles at $\bm{u_{0,N}}$.  We have derived expansions at 
these points, given in the local parameter $w_{1,N}$, and ensured that they match.  However, we should also explicitly check what happens at the point $\bu=\bm{0}$, since $w_{1,N}$ is not a suitable local parameter here.  We can instead use the Taylor series expansion of $\sigma(\bu)$ presented in \cite{MYPAPER} and described in Section \ref{SEC_SIG}.  We differentiate this to give expansions for the $\sigma$-derivatives and then, since we are at the origin, replace the variables $u_1,\dots,u_6$ with their expansions, (\ref{uxi}), in the local parameter $\xi$.

Now, the sigma expansion was given as a sum of polynomials with increasing weight in $\bu$ and hence the expansions will have increasing order of $\xi$.  Since the functions we consider all contain ratios of $\sigma$-derivatives we will only need the leading terms from each expansion, in order to check regularity. Hence we only require a minimum amount of the sigma expansion, sufficient to give non-zero expansions for the derivatives we consider.  We find that for the functions used here, we can truncate the expansion after $C_{35}$.  

Substituting these expansions into $\varphi_2(\bu)$, we find
\begin{equation} \label{phi2_lim}
\lim_{\bu \to \bm{0}} \varphi_2(\bu) = \lim_{\xi \to 0} \left[ \frac{3}{4}\frac{\mu_1}{\mu_0}\xi^4 + \xi^8 + O(\xi^{36}) \right] = 0.
\end{equation}
So $\varphi_2(\bu)$ is regular at the origin, and hence we must ensure that $\Psi(\bu)$ is as well.  Upon substitution into $\Psi(\bu)$, we find that we must set $\eta_{22}=0$ for the expansion to be regular.  This leaves us with
\[
\Psi(\bu) = - \frac{1}{4}\frac{1}{\mu_0}\frac{\sigma_{236}(\bu)}{\sigma_{23}(\bu)},
\]
with
\begin{equation} \label{psi_lim}
\lim_{\bu \to \bm{0}} \Psi(\bu) = \lim_{\xi \to 0} \left[ 
-\frac{1}{28}\frac{\mu_3}{\mu_0}\xi^7 + \frac{1}{176}\frac{(-8\mu_2 
+ 3\mu_4\mu_3)}{\mu_0}\xi^{11} + O(\xi^{15})
\right] = 0.
\end{equation}

Now all that remains is to check the periodicity properties of the functions.  Recall equation (\ref{quas}) which gave the quasi-periodicity property of $\sigma(\bu)$.  We can differentiate and use the relations in Appendix \ref{App_RelThet1} to show that 
\begin{align*}
\sigma_{23}(\bu+\ell) &= \chi(\ell)\exp(L(\bu + \textstyle \frac{\ell}{2}, \ell))\sigma_{23}(\bu), \\
\sigma_{34}(\bu+\ell) &= \chi(\ell)\exp(L(\bu + \textstyle \frac{\ell}{2}, \ell))\sigma_{34}(\bu), \\
\sigma_{236}(\bu+\ell) &= \chi(\ell)\exp(L(\bu + \textstyle \frac{\ell}{2}, \ell)) \cdot \big[
\textstyle \frac{\partial}{du_6} L(\bu + \frac{\ell}{2}, \ell) \cdot \sigma_{23} + \sigma_{236} \big].
\end{align*}
Hence, from equation (\ref{varphi2}) we can see that $\varphi_2(\bu + \ell) = \varphi_2(\bu)$ and so is Abelian.  Similarly we can see that 
\begin{align*}
\Psi(\bu + \ell) = - \frac{1}{4}\frac{1}{\mu_0}\frac{\sigma_{236}(\bu+\ell)}{\sigma_{23}(\bu+\ell)} 
= \Psi(\bu)- \frac{1}{4}\frac{1}{\mu_0}\frac{\partial}{du_6} \big( L(\bu + \textstyle \frac{\ell}{2}, \ell) \big)
\end{align*}
Hence $D_1(\Psi(\bu))$ is Abelian, and we may now write our integrand $\varphi(t)dt$ as 
\begin{equation}
\varphi_2(\bu) du_1 + A_2du_1 + A_3du_2 + A_4du_4 = D_1(\Psi(\bu)) + \bm{B}^T \bm{du}, \label{Bdef}
\end{equation}
for some vector of constants $\bm{B}^T = (B_1,B_2,B_3,B_4,B_5,B_6)$.

\subsection{Evaluating the vector $\bm{B}$ } \label{SUBSEC_B}

We can evaluate the vector $\bm{B}$ by considering the integral of equation (\ref{Bdef}) at the point $\bm{u}=\bm{0}$.  We can use the expansions in (\ref{duxi}), (\ref{phi2_lim}) and (\ref{psi_lim}) to obtain the following series.
\begin{align*}
&0 = B_{6}\xi + \frac{B_{5}}{2}\xi^2 + \left( - \frac{A_{4}}{3} + \frac{B_{4}}{3}  \right)\xi^3 
- \frac{B_6}{20}\mu_{4} \xi^5 + \left( -\frac{B_{5}}{12}\mu_{4} + \frac{B_{3}}{6}\right)\xi^6 \\
&\quad + \frac{1}{28}\frac{1}{\mu_0}\big( 3\mu_{4}\mu_{0}A_{4} - \mu_{3} - 3\mu_{4}\mu_{0}B_{4} 
- 4A_{3}\mu_{0} + 4B_{2}\mu_{0}\big)\xi^7 \\
&\quad + \frac{B_{6}}{288}( 5\mu_{4}^2 - 8\mu_{3} )\xi^9 
+ \left( - \frac{B_3}{20}\mu_{4} - \frac{B_5}{20}\mu_{3} + \frac{3B_5}{80}\mu_{4}^2\right)\xi^{10} \\
&\quad - \frac{1}{352}\frac{1}{\mu_0}\big(32A_{2}\mu_{0} - 32B_{1}\mu_{0} + 16\mu_{2}  
- 6\mu_{4}\mu_{3} - 24\mu_{0}A_{4}\mu_{3} + 21\mu_{0}A_{4}\mu_{4}^2 \\
&\quad + 24\mu_{0}B_{4}\mu_{3} - 21\mu_{0}B_{4}\mu_{4}^2 - 24\mu_{4}\mu_{0}A_{3}
+ 24\mu_{4}\mu_{0}B_{2} \big)\xi^{11} + O(\xi^{13})
\end{align*}
Setting each coefficient of $\xi$ to zero, we find
\begin{align*}
\begin{array}{lllll}
B_1 = \displaystyle \frac{1}{2} \frac{(\mu_2 + 2A_2\mu_0)}{\mu_0}, & \quad & 
B_2 = \displaystyle \frac{1}{4}\frac{(4A_3\mu_0 + \mu_3)}{\mu_0}, & \quad & B_3 = 0, \\
B_4 = A_4, & & B_5 = 0, & & B_6 = 0.
\end{array}
\end{align*}

\section{Obtaining an explicit formula for $\lambda(p)$} \label{SEC_lambda_explicit}

We now use the results of Section \ref{SEC_EI} to derive an explicit formula for the mapping $\lambda(p)$.  Start by applying the change of coordinates given in (\ref{canonical_map}) to $\lambda(p)$ as given in (\ref{TheMapping}).
\begin{align*}
\lambda(p) &= p + \int_{\infty}^p [ \varphi(p') - 1] dp' 
=  \left(\hat{p}_8 - \frac{1}{t} \right) +  \int_{0}^{\frac{1}{\hat{p}_8 - p}} \left( \varphi(t) - \frac{1}{t^2}  \right) dt \\
&= \left(\hat{p}_8 - \frac{1}{t} \right) - \int_{0}^{\frac{1}{\hat{p}_8-p}} \left[ \frac{1}{t^2} \right] dt \\ 
&\qquad + \int_{0}^{\frac{1}{\hat{p}_8-p}} \left( K[ A_4t^4+A_3t^3+A_2t^2+A_1t+1]\frac{1}{t^2}\frac{dt}{4s^3} \right),
\end{align*}
where the constants $A_1,A_2,A_3,A_4$ and $K$ were defined by equation(\ref{integrand2}).  Note from (\ref{t_second}) that
\begin{equation}
p = \hat{p}_8 + \frac{\sigma_{34}(\bu)}{\sigma_{23}(\bu)}, \qquad \bu \in \Theta^{[1]}. \label{p_T1}
\end{equation}
So let us take $\bu \in \Theta^{[1]}$, and use (\ref{p_T1}) and the evaluation of $\varphi(t)dt$ from the previous section to write $\lambda(p)$ as
\begin{align*}
&\lambda(p) = \left(\hat{p}_8 + \frac{\sigma_{34}(\bu)}{\sigma_{23}(\bu)} \right)
- \int_{0}^{\frac{1}{\hat{p}_8-p}} \left[ \frac{1}{t^2} \right] dt   
+ K \int_{0}^{\frac{1}{\hat{p}_8 - p}} \Bigg[ \frac{1}{2}\frac{(\mu_2 + 2A_2\mu_0)}{\mu_0}du_1 \\
&+ \frac{1}{4}\frac{(4A_3\mu_0 + \mu_3)}{\mu_0}du_2 + A_4du_4  \Bigg] 
+ K \int_{0}^{\frac{1}{\hat{p}_8 - p}} \left[ \frac{d}{du_1}\left( 
- \frac{1}{4}\frac{1}{\mu_0}\frac{\sigma_{236}(\bu)}{\sigma_{23}(\bu)} \right) \right] du_1.
\end{align*}
Integrating, gives
\begin{align*}
&\lambda(p) = \left(\hat{p}_8 + \frac{\sigma_{34}(\bu)}{\sigma_{23}(\bu)} \right) 
- \left[ \frac{\sigma_{34}(\bu)}{\sigma_{23}(\bu)} \right]  
+ K \Bigg[ \frac{1}{2}\frac{(\mu_2 + 2A_2\mu_0)}{\mu_0}u_1 \\
&\quad + \frac{1}{4}\frac{(4A_3\mu_0 + \mu_3)}{\mu_0}u_2 + A_4u_4  \Bigg] 
+ K \left[ - \frac{1}{4}\frac{1}{\mu_0}\frac{\sigma_{236}(\bu)}{\sigma_{23}(\bu)} \right] + \hat{C} \\
&= \hat{p}_8  + K \left[ - \frac{1}{4\mu_0}\frac{\sigma_{236}(\bu)}{\sigma_{23}(\bu)} 
+ \frac{\mu_2 + 2A_2\mu_0}{2\mu_0}u_1
+ \frac{4A_3\mu_0 + \mu_3}{4\mu_0}u_2 + A_4u_4  \right] + \hat{C}.
\end{align*}
for some constant $\hat{C}$.  We can determine $\hat{C}$ by ensuring that the following condition on the mapping is satisfied.
\[
\lim_{p \to \infty} \lambda(p) = p + O\left(\frac{1}{p}\right)
\]
Note from (\ref{p_T1}) that $p \to \infty$ implies $\sigma_{23}(\bu) \to 0$ and therefore $\bu \to \bm{u_{0,N}}$.  
\begin{align*}
&\lim_{p \to \infty} [\lambda(p)-p] = \lim_{\bu \to \bm{u_{0,N}}} \Bigg[ \hat{C} 
- \frac{\sigma_{34}(\bu)}{\sigma_{23}(\bu)} \\
&\qquad + K \Bigg( - \frac{1}{4}\frac{1}{\mu_0}\frac{\sigma_{236}(\bu)}{\sigma_{23}(\bu)} 
+ \frac{\mu_2 + 2A_2\mu_0}{2\mu_0}u_1
+ \frac{4A_3\mu_0 + \mu_3}{4\mu_0}u_2 + A_4u_4  \Bigg) \Bigg]
\end{align*}
Let us ensure that the condition is met on the sheet of the surface $C$ associated with $\lim_{t\to0} (s) = \mu_0^{1/4}$.  We can write this as a series expansion in the local parameter $w_{1}$, (as described in Section \ref{SEC_phi2_u0}).  Recall that $u_i = w_i + u_{0,i}$, and use the expansions (\ref{wi_w1}) and the  existing expansions for the $\sigma$-derivatives to obtain 
\begin{align*}
&\lim_{p \to \infty} [\lambda(p)-p] = 
\left[\frac{1}{4}\frac{1}{\mu_0^{\frac{3}{4}}} - \frac{K}{16}\frac{1}{\mu_0^{\frac{3}{2}}}\right]\frac{1}{w_1} 
+ \Bigg[ \hat{C} - \frac{3}{8}\frac{\mu_1}{\mu_0} + K \Bigg( 
- \frac{1}{4}\frac{1}{\mu_0}\frac{\sigma_{226}(\bm{u_0})}{\sigma_{22}(\bm{u_0})} \\
&+ \frac{1}{32}\frac{\mu_1}{\mu_0^{7/4}} + \frac{(\mu_2 + 2A_2\mu_0)}{2\mu_0}u_{0,1}   
+ \frac{(4A_3\mu_0 + \mu_3)}{4\mu_0}u_{0,2} + A_4u_{0,4}  \Bigg)  \Bigg] + O(w_1).
\end{align*}
Therefore, we must set the constants of integration, $\hat{C}$, to be
\begin{align*}
\hat{C} &= \frac{3}{8}\frac{\mu_1}{\mu_0} + K \Bigg( 
+ \frac{1}{4}\frac{1}{\mu_0}\frac{\sigma_{226}(\bm{u_0})}{\sigma_{22}(\bm{u_0})} 
- \frac{1}{32}\frac{\mu_1}{\mu_0^{7/4}} - \frac{1}{2}\frac{(\mu_2 + 2A_2\mu_0)}{\mu_0}u_{0,1} \\  
&\quad - \frac{1}{4}\frac{(4A_3\mu_0 + \mu_3)}{\mu_0}u_{0,2} - A_4u_{0,4}  \Bigg)  
\end{align*}
giving us the following explicit formula for the mapping $\lambda(p)$.
\begin{align*}
&\lambda(p) = \hat{p}_8 + \frac{3}{8}\frac{\mu_1}{\mu_0} + K \Bigg[ - \frac{1}{4}\frac{1}{\mu_0} \Bigg[ \frac{\sigma_{236}(\bu)}{\sigma_{23}(\bu)} - \frac{\sigma_{226}(\bm{u_0})}{\sigma_{22}(\bm{u_0})} \Bigg]
- \frac{1}{32}\frac{\mu_1}{\mu_0^{7/4}} \\
&\quad + \frac{\mu_2 + 2A_2\mu_0}{2\mu_0}(u_1 - u_{1,0})
+ \frac{4A_3\mu_0 + \mu_3}{4\mu_0}(u_2-u_{2,0}) + A_4(u_4 - u_{4,0})  \Bigg],
\end{align*}
where $\bu \in \Theta^{[1]}$ and $\bm{u_0}$ is the point on the principle sheet of the surface $C$ where $t=0$.

\section*{Acknowledgements}

We would like to thank Prof.~C.~Eilbeck and Dr.~V.~Enolski for useful conversations.

\newpage

\appendix

\section{Deriving defining relations for the strata from Jorgenson's Theorem} \label{App_Jorg}

Consider Theorem \ref{Jorgenson} in the case when $k=5$.  Then for arbitrary $\bm{a},\bm{b}$
\[
\frac{ \sum_{j=1}^6 a_j \sigma_j(\bu) }{ \sum_{j=1}^6 b_j \sigma_j(\bu) }
= \frac{\det \big[ \bm{a} \big| \bm{du}(P_1) \big| \cdots \big| \bm{du}(P_5) \big] } 
{       \det \big[ \bm{b} \big| \bm{du}(P_1) \big| \cdots \big| \bm{du}(P_5)  \big] }.
\]
Now, as $\bu \in \Theta^{[5]}$ approaches $\Theta^{[4]}$ we have the point $P_5 = (t_5,s_5)$ approaching $\infty$.  We can use the local coordinate $\xi$ here and hence replace the final column of the determinants by the expansions (\ref{duxi}).  When $\bu$ arrives at $\Theta^{[4]}$ we will have $\xi=0$ and hence the determinant in the numerator becomes
\[
\left| \begin{array}{ccccc}
a_1 & \frac{dt_1}{4s_1^3}       & \cdots & \frac{dt_4}{4s_4^3}       & 0   \\
a_2 & \frac{t_1dt_1}{4s_1^3}    & \cdots & \frac{t_4dt_4}{4s_4^3}    & 0   \\
a_3 & \frac{s_1dt_1}{4s_1^3}    & \cdots & \frac{s_4dt_4}{4s_4^3}    & 0   \\
a_4 & \frac{t_1^2dt_1}{4s_1^3}  & \cdots & \frac{t_4^2dt_4}{4s_4^3}  & 0   \\
a_5 & \frac{t_1s_1dt_1}{4s_1^3} & \cdots & \frac{t_4s_4dt_4}{4s_4^3} & 0   \\
a_6 & \frac{s_1^2dt_1}{4s_1^3}  & \cdots & \frac{s_4^2dt_4}{4s_4^3}  & -1
\end{array} \right| 
= \frac{dt_1}{3s_1^3} \cdots \frac{dt_4}{3s_4^3}
\left| \begin{array}{ccccc}
a_1 & 1      & \cdots & 1          & 0   \\
a_2 & t_1    & \cdots & t_4    & 0   \\
a_3 & s_1    & \cdots & s_4    & 0   \\
a_4 & t_1^2  & \cdots & t_4^2  & 0   \\
a_5 & t_1s_1 & \cdots & t_4s_4 & 0   \\
a_6 & s_1^2  & \cdots & s_4^2  & -1
\end{array} \right|
\]
The determinant in the denominator will be identical, except with the entries of $\bm{a}$ replaced by the entries of $\bm{b}$.  Hence the factored terms will cancel,  leaving us with the simpler determinants.  It is clear from the final column, that when we expand the determinants the resulting quotient of polynomials will not vary with the arbitrary constant $a_6$.  Hence we must conclude that for $\bu \in \Theta^{[4]}$, $\sigma_6(\bu)=0$.  The same conclusion could have been drawn from considering $b_6$.
\[
\Theta^{[4]} = \{ \bu \hspace*{0.05in} | \hspace*{0.05in} \sigma(\bu) = \sigma_6(\bu) = 0 \}.
\]
We repeat this process by considering Theorem \ref{Jorgenson} in the case when $k=4$.
\[
\frac{ \sum_{j=1}^6 a_j \sigma_j(\bu) }{ \sum_{j=1}^6 b_j \sigma_j(\bu) }
= \frac{\det \big[ \bm{a} \big| \bm{du}(P_1) \big| \cdots \big| \bm{du}(P_4) \big| \bm{du}(P_4)^{(1)} \big] } 
{       \det \big[ \bm{b} \big| \bm{du}(P_1) \big| \cdots \big| \bm{du}(P_4) \big| \bm{du}(P_4)^{(1)} \big] }.
\]
This time we consider $\bu$ descending to $\Theta^{[3]}$, by letting the fourth point move towards infinity.  The penultimate column in each determinant can be given with the expansions (\ref{duxi}) as before.  For the final column we will need to determine the derivative of these expansions:
\begin{align}
\frac{d^2u_1}{d\xi^2} = -10 \xi^{9} + O(\xi^{10})  
\qquad \frac{d^2u_4^2}{d\xi} &= \textstyle -2\xi  + \frac{9}{2}\mu_4\xi^{5} + O(\xi^6) \nonumber \\
\frac{d^2u_2}{d\xi^2} = -6\xi^{5} + O(\xi^6)  \hspace*{0.1in}    
\qquad \frac{d^2u_5}{d\xi^2} &= \textstyle -1 + \frac{5}{2}\mu_4\xi^4 + O(\xi^5) \label{dduxi} \\
\frac{d^2u_3}{d\xi^2} = -5\xi^{4} + O(\xi^5)  \hspace*{0.1in}
\qquad \frac{d^2u_6}{d\xi^2} &= \textstyle \mu_4\xi^3 + O(\xi^4). \nonumber
\end{align}
When $\bu$ arrives at $\Theta^{[3]}$ we will have $\xi=0$.  Our determinants will again factor and cancel to leave the numerator as 
\begin{align*}
&\left| \begin{array}{cccccc}
a_1 & 1      & \cdots & 1      & 0  & 0  \\
a_2 & t_1    & \cdots & t_3    & 0  & 0  \\
a_3 & s_1    & \cdots & s_3    & 0  & 0  \\
a_4 & t_1^2  & \cdots & t_3^2  & 0  & 0  \\
a_5 & t_1s_1 & \cdots & t_3s_3 & 0  & -1  \\
a_6 & s_1^2  & \cdots & s_3^2  & -1 & 0
\end{array} \right| 
\end{align*}
with the denominator identical with $\bm{b}$ instead of $\bm{a}$.  From the final two columns it is clear that the resulting quotient of polynomials will not vary with the arbitrary constant $a_6$ and $a_5$.  Hence we must conclude that 
\[
\Theta^{[3]} = \{ \bu \hspace*{0.05in} | \hspace*{0.05in} \sigma(\bu) = \sigma_6(\bu) = \sigma_5(\bu) = 0 \}.
\]
We repeat the procedure once more for $k=3$.  We let $\bu$ descend to $\Theta^{[2]}$ and use the previous expansions along with the derivatives of (\ref{dduxi}) for the final three columns.  We let $\xi=0$, cancel the common factors and expand the determinants to reduce the statement to 
\begin{equation} \label{Jorgenson_t_2}
\frac{ \sum_{j=1}^6 a_j \sigma_j(\bu) }{\sum_{j=1}^6 b_j \sigma_j(\bu)} 
= {\frac {a_{{1}}{\it t_1}\,{\it s_2}-a_{{1}}{\it s_1}\,{\it t_2}+a_
{{2}}{\it s_1}-a_{{2}}{\it s_2}-a_{{3}}{\it t_1}+a_{{3}}{\it t_2}}
{b_{{1}}{\it t_1}\,{\it s_2}-b_{{1}}{\it s_1}\,{\it t_2}+b_{{2}}{
\it s_1}-b_{{2}}{\it s_2}-b_{{3}}{\it t_1}+b_{{3}}{\it t_2}}}
\end{equation}
for $\bu \in \Theta^{[2]}$.  We therefore conclude that 
\[
\Theta^{[2]} = \{ \bu \hspace*{0.05in} | \hspace*{0.05in} \sigma(\bu) = \sigma_6(\bu) = \sigma_5(\bu) = \sigma_4(\bu) = 0 \}.
\]
Finally we consider Theorem \ref{Jorgenson} in the case when $k=2$.  Here, when we let $\bu$ descend to $\Theta^{[1]}$, we find that the statement of the Theorem involves singular matrices, (resulting from the final set of series for the derivatives all equalling zero when $\xi=0$), and hence gives us no information.

Instead we can consider equation (\ref{Jorgenson_t_2}) which held for $\bu \in \Theta^{[2]}$.  Let $\bu$ descend to $\Theta^{[1]}$ here, by using the expansions in $\xi$ for $(t_2,s_2)$.  We find that  
\begin{equation} \label{Jorgenson_t_1}
\frac{ \sum_{j=1}^6 a_j \sigma_j(\bu) }{\sum_{j=1}^6 b_j \sigma_j(\bu)} 
= \frac{a_1t_1 -a_2}{b_1t_1 - b_2} + O(\xi)
\end{equation}
and so for $u \in \Theta^{[1]}$ we can see there is no dependence on $a_3,a_4,a_5$ or $a_6$.  Hence
\[
\Theta^{[1]} = \{ \bu \hspace*{0.05in} | \hspace*{0.05in} \sigma(\bu) = \sigma_6(\bu) = \sigma_5(\bu) = \sigma_4(\bu) = \sigma_3(\bu) = 0 \}.
\]

\newpage

\section{Relations between $\bm{\sigma}$-derivatives on $\bm{\Theta^{[1]}}$} \label{App_RelThet1}

The following list of equations are valid for $\bu \in \Theta^{[1]}$.  This set contains all those relations we have obtained that express $n$-index $\sigma$-functions for $n \leq 4$.  A larger set that includes relations for $n>4$ is available online at \cite{Bweb}.
\begin{align*}
&\begin{array}{lllll}
\sigma_{1}=0 & \quad &\sigma_{11}=0,                                   & \sigma_{24}=0            & \sigma_{44}=0 \\
\sigma_{2}=0 &       &\sigma_{12}=0,                                   & \sigma_{25}=-\sigma_{34} & \sigma_{45}=0 \\
\sigma_{3}=0 &       &\sigma_{13}=0,                                   & \sigma_{26}=0            & \sigma_{46}=0 \\
\sigma_{4}=0 &       &\sigma_{14}= \textstyle -\frac{1}{2}\sigma_{22}, & \sigma_{33}=0            & \sigma_{55}=0 \\
\sigma_{5}=0 &       &\sigma_{15}= -\sigma_{23},                       & \sigma_{35}=0            & \sigma_{56}=0 \\ 
\sigma_{6}=0 &       &\sigma_{16}=0,                                   & \sigma_{36}=0            & \sigma_{66}=0
\end{array} \\ 
&\begin{array}{lll}
\sigma_{{112}}=\mu_{0}\sigma_{{34}} + \mu_{{1}} \sigma_{{23}}
& \sigma_{{156}}=-\sigma_{{236}}
& \sigma_{{445}}=0 \\
\sigma_{{113}}=0 
& \sigma_{{166}}=\sigma_{{23}}
& \sigma_{{446}}=0 \\
\sigma_{{114}}=-\sigma_{{122}} + \mu_{{1}}\sigma_{{34}} + \mu_{{2}}\sigma_{{23}} 
& \sigma_{{244}}=\sigma_{{23}} + \mu_{{4}}\sigma_{{34}}
& \sigma_{{455}}=0 \\
\sigma_{{115}}=-2\sigma_{{123}} 
& \sigma_{{245}}=-\frac{1}{2}\sigma_{{344}}
& \sigma_{{456}}=0 \\
\sigma_{{116}}=0 
& \sigma_{{246}}=0
& \sigma_{{466}}=0  \\
\sigma_{{124}}=-\frac{1}{6}\sigma_{{222}} + \frac{1}{2}\mu_{{2}}\sigma_{{34}} 
+ \frac{1}{2}\mu_{{3}}\sigma_{{23}} 
& \sigma_{{255}}=-2\sigma_{{345}}
& \sigma_{{555}}=0 \\
\sigma_{{125}}=-\frac{1}{2}\sigma_{{223}} - \sigma_{{134}} 
& \sigma_{{256}}=-\sigma_{{346}}
& \sigma_{{556}}=0 \\
\sigma_{{126}}=0 
& \sigma_{{266}}=\sigma_{{34}}
& \sigma_{{566}}=0 \\
\sigma_{{133}}=0 
& \sigma_{{333}}=0
& \sigma_{{666}}=0 \\
\sigma_{{135}}=-\frac{1}{2}\sigma_{{233}} 
& \sigma_{{335}}=0
&  \\
\sigma_{{136}}=0 
& \sigma_{{336}}=-2\sigma_{{23}}
&  \\
\sigma_{{144}}=-\sigma_{{224}} + \mu_{{4}}\sigma_{{23}} + \mu_{{3}}\sigma_{{34}}
& \sigma_{{355}}=0
& \\
\sigma_{{145}}=-\sigma_{{234}} - \frac{1}{2}\sigma_{{225}} 
& \sigma_{{356}}=-\sigma_{{34}}
& \\
\sigma_{{146}}=-\frac{1}{2}\sigma_{{226}}
& \sigma_{{366}}=0
& \\
\sigma_{{155}}=-\sigma_{{334}} - 2\sigma_{{235}}
& \sigma_{{444}}=3\sigma_{{34}}
&
\end{array} \\
&\begin{array}{ll}
\sigma_{{1136}}=-\mu_{{0}}\sigma_{{34}} \\
\sigma_{{1144}}=\mu_{{3}}\sigma_{{223}} - \frac{1}{6}\sigma_{{2222}} 
+ 2( \mu_{{4}}\sigma_{{123}} + \mu_{{2}}\sigma_{{234}} + \mu_{{3}}\sigma_{{134}} - \sigma_{{1224}}  )
+ \mu_{{1}}\sigma_{{344}}  \\
\sigma_{{1145}}=\mu_{{2}}\sigma_{{235}} -2\sigma_{{1234}} - \sigma_{{1225}} - \frac{1}{3}\sigma_{{2223}} 
+ \frac{1}{2} \mu_{{3}}\sigma_{{233}} + \mu_{{2}}\sigma_{{334}} + \mu_{{1}}\sigma_{{345}}  \\
\sigma_{{1146}}=-\sigma_{{1226}} + \mu_{{2}}\sigma_{{22}} 
+ \mu_{{2}}\sigma_{{236}} + \mu_{{1}}\sigma_{{346}} \\
\sigma_{{1155}}=-2\sigma_{{1334}}-4\sigma_{{1235}}-\sigma_{{2233}}+2\mu_{{2}}\sigma_{{22}} \\
\sigma_{{1156}}=-\mu_{{1}}\sigma_{{34}} - 2\sigma_{{1236}} \\
\sigma_{{1166}}=2\sigma_{{123}} \\
\sigma_{{1244}}=-\frac{1}{3}\sigma_{{2224}}+\sigma_{{123}} + \frac{1}{2}\mu_{{4}}\sigma_{{223}}
+\mu_{{4}}\sigma_{{134}} + \frac{1}{2}\mu_{{2}}\sigma_{{344}} + \mu_{{3}}\sigma_{{234}} \\
\sigma_{{1245}}= \frac{1}{4}\mu_{{4}}\sigma_{{233}} -\frac{1}{6}\sigma_{{2225}} 
- \frac{1}{2}\sigma_{{2234}} - \frac{1}{2}\sigma_{{1344}} + \frac{1}{2}\mu_{{3}}\sigma_{{334}} \\
\hspace*{0.7in} + \frac{1}{2}\mu_{{2}}\sigma_{{345}} + \frac{1}{2}\mu_{{3}}\sigma_{{235}} \\
\end{array}
\end{align*}
\begin{align*}
&\begin{array}{ll}
\sigma_{{1246}}=\frac{1}{2}\mu_{{3}}\sigma_{{22}} - \frac{1}{6}\sigma_{{2226}}
+\frac{1}{2}\mu_{{3}}\sigma_{{236}} + \frac{1}{2}\mu_{{2}}\sigma_{{346}} \\
\sigma_{{1255}}=\mu_{{3}}\sigma_{{22}} - \sigma_{{2235}} - 2\sigma_{{1345}} - \sigma_{{2334}} \\
\sigma_{{1256}}=-\frac{1}{2}\sigma_{{2236}} - \frac{1}{2}\mu_{{2}}\sigma_{{34}} - \sigma_{{1346}}  \\
\sigma_{{1266}}=\frac{1}{2}\sigma_{{223}} + \sigma_{{134}} \\
\sigma_{{1333}}=-6\mu_{{0}}\sigma_{{34}} + 2\mu_{{1}}\sigma_{{23}} \\
\sigma_{{1335}}=\frac{4}{3}\mu_{{2}}\sigma_{{23}} - \frac{4}{3}\mu_{{1}}\sigma_{{34}}
- \frac{1}{3}\sigma_{{2333}} \\
\sigma_{{1336}}= - 2\sigma_{{123}} \\
\sigma_{{1355}}= - \sigma_{{2335}} - \frac{1}{3}\sigma_{{3334}} - \frac{2}{3}\mu_{{2}}\sigma_{{34}}
+ 2\mu_{{3}}\sigma_{{23}} \\
\sigma_{{1356}}= - \frac{1}{2}\sigma_{{223}} - \sigma_{{134}} - \frac{1}{2}\sigma_{{2336}} \\
\sigma_{{1366}}=\frac{1}{2}\sigma_{{233}} \\
\sigma_{{1444}}=\frac{3}{2}\sigma_{{223}} + 3\sigma_{{134}} - \frac{3}{2}\sigma_{{2244}} + 3\mu_{{4}}\sigma_{{234}} 
+ \frac{3}{2}\mu_{{3}}\sigma_{{344}} \\
\sigma_{{1445}}=\mu_{{3}}\sigma_{{345}} + \frac{1}{2}\sigma_{{233}} - \sigma_{{2245}} 
- \sigma_{{2344}} + \mu_{{4}}\sigma_{{334}} + \mu_{{4}}\sigma_{{235}} \\
\sigma_{{1446}}=\mu_{{4}}\sigma_{{22}} - \sigma_{{2246}} + \mu_{{3}}\sigma_{{346}} 
+ \mu_{{4}}\sigma_{{236}} \\
\sigma_{{1455}}= - \frac{1}{2}\sigma_{{3344}}-2\sigma_{{2345}}-\frac{1}{2}\sigma_{{2255}} 
+ \mu_{{4}}\sigma_{{22}} \\
\sigma_{{1456}}= - \frac{1}{2}\sigma_{{2256}} - \frac{1}{2}\mu_{{3}}\sigma_{{34}} - \sigma_{{2346}} \\
\sigma_{{1466}}=\sigma_{{234}} - \frac{1}{2}\sigma_{{2266}} \\
\sigma_{{1555}}= - 3\sigma_{{2355}} + 8\mu_{{4}}\sigma_{{23}} - 3\sigma_{{3345}} \\
\sigma_{{1556}}= - 2\sigma_{{234}} - \sigma_{{3346}} - 2\sigma_{{2356}}
\end{array} \\
&\begin{array}{ll}
\sigma_{{1566}}=\sigma_{{334}} - \sigma_{{2366}} + \sigma_{{235}} 
& \sigma_{{3556}}= - 2\sigma_{{345}}  \\
\sigma_{{1666}}=\sigma_{{22}} + 3\sigma_{{236}}
& \sigma_{{3566}}= - 2\sigma_{{346}} \\
\sigma_{{2444}}=3\sigma_{{234}} + \frac{3}{2}\mu_{{4}}\sigma_{{344}} 
&  \sigma_{{3666}}=\sigma_{{34}} \\
\sigma_{{2445}}=\sigma_{{334}} + \sigma_{{235}} - \frac{1}{3}\sigma_{{3444}} + \mu_{{4}}\sigma_{{345}} 
&  \sigma_{{4444}}=6\sigma_{{344}} \\
\sigma_{{2446}}=\sigma_{{22}} + \sigma_{{236}} + \mu_{{4}}\sigma_{{346}} 
& \sigma_{{4445}}=3\sigma_{{345}} \\
\sigma_{{2455}}=\sigma_{{22}} - \sigma_{{3445}} 
& \sigma_{{4446}}=3\sigma_{{346}} \\
\sigma_{{2456}}= - \frac{1}{2}\mu_{{4}}\sigma_{{34}} - \frac{1}{2}\sigma_{{3446}} 
& \sigma_{{4455}}=0 \\
\sigma_{{2466}}=\frac{1}{2}\sigma_{{344}} 
& \sigma_{{4456}}= - \sigma_{{34}} \\
\sigma_{{2555}}=10\sigma_{{23}} + 2\mu_{{4}}\sigma_{{34}} - 3\sigma_{{3455}}
& \sigma_{{4466}}=0 \\
\sigma_{{2556}}= - \sigma_{{344}} - 2\sigma_{{3456}} 
& \sigma_{{4556}}=0 \\
\sigma_{{2666}}=3\sigma_{{346}} 
& \sigma_{{4566}}=0 \\
\sigma_{{3333}}=0 
& \sigma_{{4666}}=0 \\
\sigma_{{3335}}=0 
&  \sigma_{{5555}}=0 \\
\sigma_{{3336}}= - 3\sigma_{{233}} 
& \sigma_{{5556}}=0 \\
\sigma_{{3355}}=0 
& \sigma_{{4555}}=4\sigma_{{34}} \\
\sigma_{{3356}}= - 2\sigma_{{334}} - 2\sigma_{{235}} 
&  \sigma_{{5566}}=0 \\
\sigma_{{3366}}= - 2\sigma_{{22}} - 4\sigma_{{236}} 
& \sigma_{{5666}}=0  \\
\sigma_{{3466}}=\sigma_{{345}} - \sigma_{{2566}} 
& \sigma_{{6666}}=0 \\
\sigma_{{3555}}=0 
\end{array}
\end{align*}

\newpage

\section{Relations between $\bm{\sigma}$-derivatives at $\bm{u_{0,N}}$} \label{APP_Rel_u0}

The following list of equations were not valid in general for $\bu \in \Theta^{[1]}$, but are true at 
$\bu = \bm{u_{0,N}}$.  This set contains all those relations we have obtained that express $n$-index $\sigma$-functions for $n \leq 4$.  A larger set that includes relations for $n>4$ is available online at \cite{Bweb}.
\begin{align*}  
\begin{array}{ll} 
\sigma_{34} = \displaystyle \frac{1}{2} \frac{ \sigma_{22} }{ {\it i}^N \mu_{0}^{1/4} } 
& \sigma_{{233}}= \displaystyle -{\it i}^N \mu_{0}^{1/4}\sigma_{{22}} \\
\sigma_{{111}}= \displaystyle 0 
& \sigma_{{234}}= \displaystyle \frac{1}{6}{\frac {\sigma_{{222}}}{{\it i}^N \mu_{0}^{1/4} }}
- \frac{1}{4}{\frac {\mu_{{2}}\sigma_{{22}}}{{{\it i}}^{2N}\sqrt {\mu_{{0}}}}} \\
\sigma_{{112}}= \displaystyle \frac{1}{2}\sigma_{{22}}{{\it i}}^{3N}{\mu_{{0}}}^{3/4}
& \sigma_{{235}}= \displaystyle - \frac{1}{2}{\frac {\sigma_{{223}}}{{\it i}^N \mu_{0}^{1/4} }}  \\
\sigma_{{113}}= \displaystyle 0 
& \sigma_{{236}}= \displaystyle -\frac{1}{2}\sigma_{{22}} \\
\sigma_{{123}}= \displaystyle -\frac{1}{2}{{\it i}}^{2N}\sqrt{\mu_{{0}}}\sigma_{{22}}
& \sigma_{{334}}= \displaystyle + {\frac {\sigma_{{223}}}{{\it i}^N \mu_{0}^{1/4}  }} \\
\sigma_{{134}}= \displaystyle \frac{ {{\it i}}^{3N}\sigma_{{122}} }{ 2\mu_{0}^{1/4}  }
- {\frac{{{\it i}}^{2N}\sigma_{{22}}\mu_{{1}}}{2\sqrt{\mu_{{0}}}}} - \frac{\sigma_{{223}}}{2} 
& \sigma_{{344}}= \displaystyle -\frac{1}{2}{\frac {\sigma_{{22}}\mu_{{3}}}{{{\it i}}^{2N}\sqrt {\mu_{{0}}}}}
+ {\frac {\sigma_{{224}}}{{\it i}^N \mu_{0}^{1/4}  }}  
\end{array} 
\end{align*}
\begin{align*}
\sigma_{{345}} &= \displaystyle {\frac {{{\it i}}^{3N}\sigma_{{225}}}{2 \mu_{0}^{1/4} }}
+ {\frac {{{\it i}}^{2N}\sigma_{{222}}}{6\sqrt {\mu_{{0}}}}}
- {\frac {{{\it i}}^{N}\mu_{{2}}\sigma_{{22}}}{4{\mu_{{0}}}^{3/4}}} \\
\sigma_{{346}}&=\displaystyle \frac{1}{2}{\frac {\sigma_{{226}}}{{\it i}^N\mu_{0}^{1/4}}} \\
\sigma_{{1111}}&= - 6{\mu_{{0}}}^{\frac{3}{2}}\sigma_{{22}}{{\it i}}^{2N} \\
\sigma_{{1112}}&=\textstyle - 3\mu_{{1}}{{\it i}}^{2N}\sqrt {\mu_{{0}}}\sigma_{{22}} 
+ \frac{3}{2} {{\mu_{{0}}}^{\frac{3}{4}}\sigma_{{122}}}{{\it i}}^{3N}\\
\sigma_{{1113}}&=\textstyle - \frac{3}{2}{\mu_{{0}}}^{5/4}\sigma_{{22}}{{\it i}}^{N}\\
\sigma_{{1114}}&= - \frac{9}{4}{{\it i}}^{2N}\sqrt {\mu_{{0}}}\mu_{{2}}\sigma_{{22}} 
- \frac{3}{2}{\frac {{{\it i}}^{2N}{\mu_{{1}}}^{2}\sigma_{{22}}}{\sqrt {\mu_{{0}}}}} \\
&\qquad - \frac{3}{2}\sigma_{{1122}} + \frac{3}{2}{\frac {\mu_{{1}}\sigma_{{122}}}{{{\it i}}^{N}\mu_{0}^{1/4}  }} 
+ \frac{1}{2} { {\mu_{{0}}}^{\frac{3}{4}} \sigma_{{222}}}  {{\it i}}^{3N}\\
\sigma_{{1115}}&=\textstyle - \frac{3}{2}{{\it i}}^{N}\sigma_{{22}}\mu_{{1}}\mu_{0}^{\frac{1}{4}} 
+ 3 {\sqrt {\mu_{{0}}}\sigma_{{122}}}{{{\it i}}^{2N}}\\
\sigma_{{1116}}&=\textstyle \frac{3}{2}\sigma_{{22}}\mu_{{0}}\\
\sigma_{{1123}}&=\textstyle \frac{1}{2}{{\it i}}^{3N}{\mu_{{0}}}^{\frac{3}{4}}\sigma_{{223}} 
- \sqrt {\mu_{{0}}}{{\it i}}^{2N}\sigma_{{122}}
\end{align*}
\begin{align*}
\sigma_{{1124}}&= - \frac{3}{4}{\frac {{{\it i}}^{2N}\mu_{{2}}\sigma_{{22}}\mu_{{1}}}{\sqrt {\mu_{{0}}}}} 
- \frac{3}{4}{{\it i}}^{2N}\sqrt {\mu_{{0}}}\sigma_{{22}}\mu_{{3}} - \frac{1}{3}\sigma_{{1222}} 
+ \frac{1}{2} {{\mu_{{0}}}^{\frac{3}{4}}\sigma_{{224}}}{{\it i}}^{3N}  \\
&\quad + \frac{1}{6}{\frac {\mu_{{1}}\sigma_{{222}}}{{{\it i}}^{N}\mu_{0}^{1/4}}} 
+ \frac{1}{2}{\frac {\mu_{{2}}\sigma_{{122}}}{{{\it i}}^{N}\mu_{0}^{1/4}}} \\
\sigma_{{1125}}&= - \frac{1}{2}\mu_{{2}}\sigma_{{22}}{{\it i}}^{N}\mu_{0}^{\frac{1}{4}} 
- {\frac {{{\it i}}^{N}{\mu_{{1}}}^{2}\sigma_{{22}}}{{\mu_{{0}}}^{\frac{3}{4}}}} 
- \frac{1}{2}{\frac {\sigma_{{1122}}}{{{\it i}}^{N}\mu_{0}^{1/4}}} 
+ \frac{1}{2} {{\mu_{{0}}}^{\frac{3}{4}}\sigma_{{225}}}{{\it i}}^{3N} \\
&\quad + {\frac {\mu_{{1}}\sigma_{{122}}}{{{\it i}}^{2N}\sqrt {\mu_{{0}}}}} 
+ \frac{2}{3}{\frac {\sqrt {\mu_{{0}}}\sigma_{{222}}}{{{\it i}}^{2N}}}\\
\sigma_{{1126}}&=\textstyle\frac{1}{2}\mu_{{1}}\sigma_{{22}} 
+ \frac{1}{2}\sigma_{{226}}{{\it i}}^{3N}{\mu_{{0}}}^{\frac{3}{4}}\\
\sigma_{{1133}}&=2\sigma_{{22}}\mu_{{0}}\\
\sigma_{{1134}}&={\frac {{{\it i}}^{N}{\mu_{{1}}}^{2}\sigma_{{22}}}{{\mu_{{0}}}^{\frac{3}{4}}}} 
- \frac{1}{4}\mu_{{2}}\sigma_{{22}}{{\it i}}^{N}\mu_{0}^{\frac{1}{4}} 
- \sigma_{{1223}} + \frac{1}{2}{\frac {\sigma_{{223}}\mu_{{1}}}{{{\it i}}^{N}\mu_{0}^{1/4}}} 
+ \frac{1}{2}{\frac {\sigma_{{1122}}}{{{\it i}}^{N}\mu_{0}^{1/4}}} \\
&\quad - \frac{1}{2}{\frac {\sqrt {\mu_{{0}}}\sigma_{{222}}}{{{\it i}}^{2N}}} 
- {\frac {\mu_{{1}}\sigma_{{122}}}{{{\it i}}^{2N}\sqrt {\mu_{{0}}}}}\\
\sigma_{{1135}}&={{\it i}}^{2N}\sqrt {\mu_{{0}}}\sigma_{{223}}
+ {{\it i}}^{N}\mu_{0}^{\frac{1}{4}}\sigma_{{122}}\\
\sigma_{{1233}}&=\sigma_{{22}}\mu_{{1}} - {\frac {\sqrt {\mu_{{0}}}\sigma_{{223}}}{{{\it i}}^{2N}}} 
- {\frac {\sigma_{{122}}\mu_{0}^{\frac{1}{4}}}{{{\it i}}^{3N}}}\\
\sigma_{{1234}}&= - \frac{1}{2}{{\it i}}^{N}\mu_{0}^{\frac{1}{4}}\sigma_{{22}}\mu_{{3}} 
+ \frac{1}{2}{\frac {\mu_{{2}}\sigma_{{22}}{{\it i}}^{N}\mu_{{1}}}{{\mu_{{0}}}^{\frac{3}{4}}}} 
- \frac{1}{6}\sigma_{{2223}} 
+ \frac{1}{4}{\frac {\sigma_{{223}}\mu_{{2}}}{{{\it i}}^{N}\mu_{0}^{1/4}}} 
+ \frac{1}{6}{\frac {\sigma_{{1222}}}{{{\it i}}^{N}\mu_{0}^{1/4}}} \\
&\quad - \frac{1}{2}{\frac {\sqrt {\mu_{{0}}}\sigma_{{224}}}{{{\it i}}^{2N}}} 
- \frac{1}{6}{\frac {\mu_{{1}}\sigma_{{222}}}{{{\it i}}^{2N}\sqrt {\mu_{{0}}}}} 
- \frac{1}{4}{\frac {\mu_{{2}}\sigma_{{122}}}{{{\it i}}^{2N}\sqrt {\mu_{{0}}}}}\\
\sigma_{{1235}}&=\frac{1}{2}{\frac {{{\it i}}^{2N}\sigma_{{223}}\mu_{{1}}}{\sqrt {\mu_{{0}}}}} 
- \frac{1}{2}\sqrt {\mu_{{0}}}{{\it i}}^{2N}\sigma_{{225}} 
- \frac{1}{6}\mu_{0}^{\frac{1}{4}}{{\it i}}^{N}\sigma_{{222}} 
+ \frac{1}{2}\mu_{{2}}\sigma_{{22}} 
- \frac{1}{2}{\frac {\sigma_{{1223}}}{{{\it i}}^{N}\mu_{0}^{1/4}}}\\
\sigma_{{1236}}&= - \frac{1}{2}{{\it i}}^{2N}\sqrt {\mu_{{0}}}\sigma_{{226}} 
- \frac{1}{2}\sigma_{{122}}\\
\sigma_{{1334}}&= - {\frac {{{\it i}}^{2N}\sigma_{{223}}\mu_{{1}}}{\sqrt {\mu_{{0}}}}} 
+ \frac{1}{2}\mu_{{2}}\sigma_{{22}} - \frac{1}{2}\sigma_{{2233}} 
+ {\frac {\sigma_{{1223}}}{{{\it i}}^{N}\mu_{0}^{1/4}}} \\
\sigma_{{1344}}&={\frac {{{\it i}}^{N}\mu_{{3}}\sigma_{{22}}\mu_{{1}}}{{\mu_{{0}}}^{\frac{3}{4}}}} 
- \frac{3}{2}{{\it i}}^{N}\mu_{0}^{\frac{1}{4}}\sigma_{{22}}\mu_{{4}} 
+ \frac{1}{2}{\frac {{{\it i}}^{N}{\mu_{{2}}}^{2}\sigma_{{22}}}{{\mu_{{0}}}^{\frac{3}{4}}}} 
- \sigma_{{2234}} 
+ \frac{1}{6}{\frac {\sigma_{{2222}}}{{{\it i}}^{N}\mu_{0}^{1/4}}} \\
&\quad + \frac{1}{2}{\frac {\sigma_{{223}}\mu_{{3}}}{{{\it i}}^{N}\mu_{0}^{1/4}}} 
+ {\frac {\sigma_{{1224}}}{{{\it i}}^{N}\mu_{0}^{1/4}}} 
- \frac{1}{2}{\frac {\mu_{{3}}\sigma_{{122}}}{{{\it i}}^{2N}\sqrt {\mu_{{0}}}}} 
- {\frac {\mu_{{1}}\sigma_{{224}}}{{{\it i}}^{2N}\sqrt {\mu_{{0}}}}} 
- \frac{1}{3}{\frac {\mu_{{2}}\sigma_{{222}}}{{{\it i}}^{2N}\sqrt {\mu_{{0}}}}} 
\end{align*}
\begin{align*}
\sigma_{{1345}}&= - \frac{1}{2}\sigma_{{2235}} - \frac{1}{4}\sigma_{{22}}\mu_{{3}} 
+ \frac{3}{4}{\frac {\mu_{{1}}\mu_{{2}}\sigma_{{22}}}{\mu_{{0}}}} 
+ \frac{1}{2}{\frac {\sigma_{{1225}}}{{{\it i}}^{N}\mu_{0}^{1/4}}} 
- \frac{1}{6}{\frac {\sigma_{{2223}}}{{{\it i}}^{N}\mu_{0}^{1/4}}} 
+ \frac{1}{2}{\frac {\mu_{0}^{\frac{1}{4}}\sigma_{{224}}}{{{\it i}}^{3N}}} \\
&\quad - \frac{1}{4}{\frac {\mu_{{2}}\sigma_{{122}}}{{{\it i}}^{3N}{\mu_{{0}}}^{3/4}}} 
- \frac{1}{2}{\frac {\mu_{{1}}\sigma_{{225}}}{{{\it i}}^{2N}\sqrt {\mu_{{0}}}}} 
+ \frac{1}{4}{\frac {\sigma_{{223}}\mu_{{2}}}{{{\it i}}^{2N}\sqrt {\mu_{{0}}}}} 
+ \frac{1}{6}{\frac {\sigma_{{1222}}}{{{\it i}}^{2N}\sqrt {\mu_{{0}}}}} 
- \frac{1}{3}{\frac {\mu_{{1}}\sigma_{{222}}}{{{\it i}}^{3N}{\mu_{{0}}}^{3/4}}}  \\
\sigma_{{1346}}&= - \frac{1}{6}\sigma_{{222}} - \frac{1}{2}\sigma_{{2236}} 
- \frac{1}{4}{\frac {\mu_{{2}}\sigma_{{22}}}{{{\it i}}^{N}\mu_{0}^{1/4}}} 
+ \frac{1}{2}{\frac {\sigma_{{1226}}}{{{\it i}}^{N}\mu_{0}^{1/4}}} 
- \frac{1}{2}{\frac {\mu_{{1}}\sigma_{{226}}}{{{\it i}}^{2N}\sqrt {\mu_{{0}}}}}\\
\sigma_{{2333}}&= - 2{\frac {\mu_{{1}}\sigma_{{22}}{{\it i}}^{3N}}{\mu_{0}^{1/4}}} 
- 3{{\it i}}^{N}\mu_{0}^{1/4}\sigma_{{223}}\\
\sigma_{{2334}}&=\frac{1}{2}\sigma_{{22}}\mu_{{3}} 
+ \frac{1}{3}{\frac {\sigma_{{2223}}}{{{\it i}}^{N}\mu_{0}^{1/4}}} 
- \frac{1}{2}{\frac {\sigma_{{223}}\mu_{{2}}}{{{\it i}}^{2N}\sqrt {\mu_{{0}}}}} 
- {\frac {\mu_{0}^{1/4}\sigma_{{224}}}{{{\it i}}^{3N}}}\\
\sigma_{{2335}}&= - {{\it i}}^{N}\mu_{0}^{1/4}\sigma_{{225}}
- \frac{2}{3}\sigma_{{222}} - \frac{1}{2}{\frac {\sigma_{{2233}}}{{{\it i}}^{N}\mu_{0}^{1/4}}} 
+ {\frac {\mu_{{2}}\sigma_{{22}}}{{{\it i}}^{N}\mu_{0}^{1/4}}}\\
\sigma_{{2336}}&= - {{\it i}}^{N}\mu_{0}^{1/4}\sigma_{{226}} - \sigma_{{223}}\\
\sigma_{{2344}}&=\frac{1}{2}{\frac {\mu_{{2}}\sigma_{{22}}{{\it i}}^{N}\mu_{{3}}}{{\mu_{{0}}}^{3/4}}} 
- 2{{\it i}}^{N}\mu_{0}^{1/4}\sigma_{{22}} 
+ \frac{1}{2}{\frac {\sigma_{{223}}\mu_{{4}}}{{{\it i}}^{N}\mu_{0}^{1/4}}} 
+ \frac{1}{3}{\frac {\sigma_{{2224}}}{{{\it i}}^{N}\mu_{0}^{1/4}}} \\
&\quad - \frac{1}{6}{\frac {\sigma_{{222}}\mu_{{3}}}{{{\it i}}^{2N}\sqrt {\mu_{{0}}}}} 
- \frac{1}{2}{\frac {\sigma_{{224}}\mu_{{2}}}{{{\it i}}^{2N}\sqrt {\mu_{{0}}}}}\\
\sigma_{{2345}}&= - \frac{1}{2}\sigma_{{22}}\mu_{{4}} 
+ \frac{3}{8}{\frac {{\mu_{{2}}}^{2}\sigma_{{22}}}{\mu_{{0}}}} 
+ \frac{1}{6}{\frac {\sigma_{{2225}}}{{{\it i}}^{N}\mu_{0}^{1/4}}} 
- \frac{1}{2}{\frac {\sigma_{{2234}}}{{{\it i}}^{N}\mu_{0}^{1/4}}} 
+ \frac{1}{12}{\frac {\sigma_{{2222}}}{{{\it i}}^{2N}\sqrt {\mu_{{0}}}}} \\
&\quad - \frac{1}{4}{\frac {\sigma_{{225}}\mu_{{2}}}{{{\it i}}^{2N}\sqrt {\mu_{{0}}}}} 
+ \frac{1}{4}{\frac {\sigma_{{223}}\mu_{{3}}}{{{\it i}}^{2N}\sqrt {\mu_{{0}}}}} 
- \frac{1}{4}{\frac {\mu_{{2}}\sigma_{{222}}}{{{\it i}}^{3N}{\mu_{{0}}}^{3/4}}}\\
\sigma_{{2346}}&= - \frac{1}{2}\sigma_{{224}} + \frac{1}{6}{\frac {\sigma_{{2226}}}{{{\it i}}^{N}\mu_{0}^{1/4}}} 
- \frac{1}{4}{\frac {\sigma_{{226}}\mu_{{2}}}{{{\it i}}^{2N}\sqrt {\mu_{{0}}}}}\\
\sigma_{{2355}}&= - {\frac {\sigma_{{2235}}}{{{\it i}}^{N}\mu_{0}^{1/4}}} 
+ {\frac {\sigma_{{22}}\mu_{{3}}}{{{\it i}}^{N}\mu_{0}^{1/4}}} 
- \frac{1}{3}{\frac {\sigma_{{2223}}}{{{\it i}}^{2N}\sqrt {\mu_{{0}}}}} 
+ \frac{1}{2}{\frac {\sigma_{{223}}\mu_{{2}}}{{{\it i}}^{3N}{\mu_{{0}}}^{3/4}}}\\
\sigma_{{2356}}&= - \frac{1}{2}\sigma_{{225}} 
- \frac{1}{2}{\frac {\sigma_{{222}}}{{{\it i}}^{N}\mu_{0}^{1/4}}} 
- \frac{1}{2}{\frac {\sigma_{{2236}}}{{{\it i}}^{N}\mu_{0}^{1/4}}} 
+ \frac{1}{2}{\frac {\mu_{{2}}\sigma_{{22}}}{{{\it i}}^{2N}\sqrt {\mu_{{0}}}}}\\
\sigma_{{2366}}&= - \sigma_{{226}} 
+ \frac{1}{2}{\frac {\sigma_{{223}}}{{{\it i}}^{N}\mu_{0}^{1/4}}}\\
\sigma_{{2566}}&= - \frac{1}{2}{\frac {\mu_{{2}}\sigma_{{22}}{{\it i}}^{N}}{{\mu_{{0}}}^{3/4}}} 
+ \frac{1}{2}{\frac {\sigma_{{225}}}{{{\it i}}^{N}\mu_{0}^{1/4}}} 
- \frac{1}{2}{\frac {\sigma_{{2266}}}{{{\it i}}^{N}\mu_{0}^{1/4}}} 
+ \frac{1}{3}{\frac {\sigma_{{222}}}{{{\it i}}^{2N}\sqrt {\mu_{{0}}}}}
\end{align*}
\begin{align*}
\sigma_{{3334}}&=\sigma_{{222}} + \frac{3}{2}{\frac {\sigma_{{2233}}}{{{\it i}}^{N}\mu_{0}^{1/4}}} 
- \frac{5}{2}{\frac {\mu_{{2}}\sigma_{{22}}}{{{\it i}}^{N}\mu_{0}^{1/4}}}\\
\sigma_{{3344}}&=2\sigma_{{22}}\mu_{{4}} 
- {\frac {{\mu_{{2}}}^{2}\sigma_{{22}}}{2\mu_{{0}}}} 
+ 2{\frac {\sigma_{{2234}}}{{{\it i}}^{N}\mu_{0}^{1/4}}} 
- {\frac {\sigma_{{223}}\mu_{{3}}}{{{\it i}}^{2N}\sqrt {\mu_{{0}}}}} 
- \frac{1}{6}{\frac {\sigma_{{2222}}}{{{\it i}}^{2N}\sqrt {\mu_{{0}}}}} 
+ \frac{1}{3}{\frac {\mu_{{2}}\sigma_{{222}}}{{{\it i}}^{3N}{\mu_{{0}}}^{3/4}}}\\
\sigma_{{3345}}&={\frac {\sigma_{{2235}}}{{{\it i}}^{N}\mu_{0}^{1/4}}} 
- {\frac {\sigma_{{22}}\mu_{{3}}}{{{\it i}}^{N}\mu_{0}^{1/4}}} 
+ \frac{1}{3}{\frac {\sigma_{{2223}}}{{{\it i}}^{2N}\sqrt {\mu_{{0}}}}} 
- \frac{1}{2}{\frac {\sigma_{{223}}\mu_{{2}}}{{{\it i}}^{3N}{\mu_{{0}}}^{3/4}}}\\
\sigma_{{3346}}&=\frac{1}{3}{\frac {\sigma_{{222}}}{{{\it i}}^{N}\mu_{0}^{1/4}}} 
+ {\frac {\sigma_{{2236}}}{{{\it i}}^{N}\mu_{0}^{1/4}}}\\
\sigma_{{3444}}&=\frac{3}{4}{\frac {{{\it i}}^{N}\sigma_{{22}}{\mu_{{3}}}^{2}}{{\mu_{{0}}}^{3/4}}} 
+ 3/4{\frac {{{\it i}}^{N}\mu_{{4}}\mu_{{2}}\sigma_{{22}}}{{\mu_{{0}}}^{3/4}}} 
+ \frac{3}{2}{\frac {\sigma_{{223}}}{{{\it i}}^{N}\mu_{0}^{1/4}}} 
+ \frac{3}{2}{\frac {\sigma_{{2244}}}{{{\it i}}^{N}\mu_{0}^{1/4}}} \\
&\quad - \frac{3}{2}{\frac {\sigma_{{224}}\mu_{{3}}}{{{\it i}}^{2N}\sqrt {\mu_{{0}}}}} 
- \frac{1}{2}{\frac {\sigma_{{222}}\mu_{{4}}}{{{\it i}}^{2N}\sqrt {\mu_{{0}}}}}\\
\sigma_{{3445}}&= - \sigma_{{22}} + \frac{3}{4}{\frac {\mu_{{2}}\sigma_{{22}}\mu_{{3}}}{\mu_{{0}}}} 
+ {\frac {\sigma_{{2245}}}{{{\it i}}^{N}\mu_{0}^{1/4}}} 
- \frac{1}{2}{\frac {\sigma_{{225}}\mu_{{3}}}{{{\it i}}^{2N}\sqrt {\mu_{{0}}}}} 
+ \frac{1}{3}{\frac {\sigma_{{2224}}}{{{\it i}}^{2N}\sqrt {\mu_{{0}}}}} \\
&\quad - \frac{1}{3}{\frac {\sigma_{{222}}\mu_{{3}}}{{{\it i}}^{3N}{\mu_{{0}}}^{3/4}}} 
- \frac{1}{2}{\frac {\sigma_{{224}}\mu_{{2}}}{{{\it i}}^{3N}{\mu_{{0}}}^{3/4}}}\\
\sigma_{{3446}}&= - \frac{1}{2}{\frac {\sigma_{{22}}\mu_{{4}}}{{{\it i}}^{N}\mu_{0}^{1/4}}} 
+ {\frac {\sigma_{{2246}}}{{{\it i}}^{N}\mu_{0}^{1/4}}} 
- \frac{1}{2}{\frac {\sigma_{{226}}\mu_{{3}}}{{{\it i}}^{2N}\sqrt {\mu_{{0}}}}}\\
\sigma_{{3455}}&= - {\frac {\sigma_{{22}}\mu_{{4}}}{{{\it i}}^{N}\mu_{0}^{1/4}}} 
+ \frac{1}{2}{\frac {\sigma_{{2255}}}{{{\it i}}^{N}\mu_{0}^{1/4}}} 
+ \frac{1}{2}{\frac {{\mu_{{2}}}^{2}\sigma_{{22}}}{{{\it i}}^{N}{\mu_{{0}}}^{5/4}}} 
+ \frac{1}{3}{\frac {\sigma_{{2225}}}{{{\it i}}^{2N}\sqrt {\mu_{{0}}}}} 
- \frac{1}{2}{\frac {\sigma_{{225}}\mu_{{2}}}{{{\it i}}^{3N}{\mu_{{0}}}^{3/4}}} \\
&\quad + \frac{1}{12}{\frac {\sigma_{{2222}}}{{{\it i}}^{3N}{\mu_{{0}}}^{3/4}}} 
- \frac{1}{3}{\frac {\mu_{{2}}\sigma_{{222}}}{\mu_{{0}}}}\\
\sigma_{{3456}}&= - \frac{1}{2}{\frac {\sigma_{{224}}}{{{\it i}}^{N}\mu_{0}^{1/4}}} 
+ \frac{1}{2}{\frac {\sigma_{{2256}}}{{{\it i}}^{N}\mu_{0}^{1/4}}} 
+ \frac{1}{6}{\frac {\sigma_{{2226}}}{{{\it i}}^{2N}\sqrt {\mu_{{0}}}}} 
+ \frac{1}{4}{\frac {\sigma_{{22}}\mu_{{3}}}{{{\it i}}^{2N}\sqrt {\mu_{{0}}}}} 
- \frac{1}{4}{\frac {\sigma_{{226}}\mu_{{2}}}{{{\it i}}^{3N}{\mu_{{0}}}^{3/4}}}
\end{align*}

\begin{footnotesize}
\bibliography{GSCTR_Benney_arXiv}{}

\begin{thebibliography}{10}

\bibitem{GibTsa1}
S.P.~Tsarev J.~Gibbons.
\newblock Reductions of the {B}enney equations.
\newblock {\em Physics Letters A}, 211:19, 1996.

\bibitem{GibTsa2}
S.P.~Tsarev J.~Gibbons.
\newblock Conformal maps and reductions of the {B}enney equations.
\newblock {\em Physics Letters A}, 258, 1999.

\bibitem{yg99}
J.~Gibbons L.~Yu.
\newblock The initial value problem for reductions of the {B}enney equations.
\newblock {\em Inverse Problems}, 16:605--618, 2000.

\bibitem{bg03}
S.~Baldwin and J.~Gibbons.
\newblock Hyperelliptic reduction of the {B}enney moment equations.
\newblock {\em J. Phys. A: Math. Gen.}, 36:8393, 2003.

\bibitem{bg04}
S.~Baldwin and J.~Gibbons.
\newblock Higher genus hyperelliptic reductions of the {B}enney equations.
\newblock {\em J. Phys. A}, 37:5341--5354, 2004.

\bibitem{bg06}
S.~Baldwin and J.~Gibbons.
\newblock Genus 4 trigonal reduction of the {B}enney equations.
\newblock {\em J. Phys. A}, 39:3607--3639, 2006.

\bibitem{benney}
D.~J. Benney.
\newblock Some properties of long nonlinear waves.
\newblock {\em Studies in Applied Mathematics}, 52:45, 1973.

\bibitem{Gib1}
Gibbons J.
\newblock Collisionless boltzmann equations and integrable moment equations.
\newblock {\em Physica D}, 3:503, 1981.

\bibitem{zakh1}
V.~E. Zakharov.
\newblock On the {B}enney equations.
\newblock {\em Physica D}, 3:193, 1981.

\bibitem{kupman1}
B.A. Kupershmidt and Yu.~I. Manin.
\newblock Long-wave equation with free boundaries.
\newblock {\em Functional Analysis and its Applications}, 11:188, 1977.

\bibitem{kupman2}
B.~A. Kupershmidt and Yu.~I. Manin.
\newblock Equations of long waves with a free surface.
\newblock {\em Functional Analysis and its Applications}, 12:20, 1977.

\bibitem{Tsa1}
S.~P. Tsarev.
\newblock The geometry of {H}amiltonian systems of hydrodynamic type. {T}he
  generalized hodograph method.
\newblock {\em Math. USSR-Izv}, 37:397, 1991.

\bibitem{MYPAPER}
M.~England and J.C. Eilbeck.
\newblock Abelian functions associated with a cyclic tetragonal curve of genus
  six.
\newblock {\em J. Phys. A: Math. Theor.}, 42:09510, 2009.

\bibitem{DoublePend}
V.~Z. Enolskii, M.~Pronine, and P.H. Richter.
\newblock Double pendulum and $\theta$-divisor.
\newblock {\em J. Nonlinear Sci.}, 13:157--174, 2003.

\bibitem{HL08}
E.~Hackmann and C.~Lammerz\"ahl.
\newblock Complete analytic solution of the geodesic equation in
  {S}chwarzschild-({A}nti-)de {S}itter spacetimes.
\newblock {\em Phys. Rev. Lett.}, 100:171101, 2008.

\bibitem{AbF00}
S.~Abenda and Yu.~N. Fedorov.
\newblock On the weak {K}owalevski-{P}ainleve property for hyperelliptic
  seperable systems.
\newblock {\em Acta Appl. Math.}, 60:138--178, 2000.

\bibitem{AlF00}
M.~S. Alber and Yu.~N. Fedorov.
\newblock Wave solutions of evolution equations and {H}amiltonian flows on
  nonlinear subvarieties of generalised jacobians.
\newblock {\em J. Phys. A}, 33:8409--8425, 2000.

\bibitem{mu83}
D.~Mumford.
\newblock {\em Tata Lectures on {T}heta {I}}, volume~28 of {\em Progress in
  Mathematics}.
\newblock Birkh\"auser, Boston, 1983.

\bibitem{bel97}
V.~M. Buchstaber, V.~Z. Enolskii, and D.~V. Leykin.
\newblock Kleinian functions, hyperelliptic {J}acobians and applications.
\newblock {\em Reviews in Math. and Math. Physics}, 10:1--125, 1997.

\bibitem{Bweb}
M.~England.
\newblock
  \texttt{http://www.ma.hw.ac.uk/$\stackrel{\sim}{}$matte/{B}enney{R}eduction/%
}.

\bibitem{jorg92}
J.~Jorgenson.
\newblock On directional derivatives of the theta function along its divisor.
\newblock {\em Israel Journal of Mathematics}, 77:273--284, 1992.

\bibitem{DM}
W.~Schreiner.
\newblock http://www.risc.uni-linz.ac.at/software/distmaple/.

\bibitem{smb03}
Wolfgang Schreiner, Christian Mittermaier, and Karoly Bosa.
\newblock Distributed {M}aple: {P}arallel computer algebra in networked
  environments, 2003.

\end{thebibliography}
\bibliographystyle{unsrt}
\end{footnotesize}

\end{document}